\documentclass[11pt]{article}

\usepackage{mathtools}
\usepackage{amssymb}
\usepackage{dsfont}
\usepackage{slashed}
\usepackage{fullpage}
\usepackage[colorlinks=true,linkcolor=blue,citecolor=magenta,linktocpage=true]{hyperref}
\usepackage{titlesec}
\titleformat*{\section}{\normalsize\bfseries}
\titleformat*{\subsection}{\normalsize\bfseries}
\titleformat*{\subsubsection}{\normalsize\bfseries}
\usepackage{cite}
\usepackage{amsmath,amssymb,braket,tikz}
\usetikzlibrary{tikzmark,calc}

\usepackage[english]{babel}
\usepackage{csquotes}
\MakeOuterQuote{"}
\DeclareMathAlphabet{\bbvar}{U}{BOONDOX-ds}{m}{n}

\makeatletter
\renewcommand{\@dotsep}{10000}
\makeatother

\usepackage[makeroom]{cancel}

\def\be{\begin{equation}}
	\def\ee{\end{equation}}
\def\bea{\begin{eqnarray}}
	\def\eea{\end{eqnarray}}
\def\beq{\begin{eqnarray}}
	\def\eeq{\end{eqnarray}}
\def\bas{\begin{subequations}\begin{eqnarray}}
		\def\eas{\end{eqnarray}\end{subequations}}
\def\nn{\nonumber}

\newcommand{\A}{{\mathcal A}}
\newcommand{\B}{{\mathcal B}}

\newcommand{\cI}{{\mathcal I}}

\newcommand{\cK}{{\mathcal K}}
\newcommand{\cL}{{\mathcal L}}

\newcommand{\cN}{{\mathcal N}}

\newcommand{\cO}{{\mathcal O}}

\newcommand{\cZ}{{\mathcal Z}}

\newcommand{\mat} [2] {\left ( \begin{array}{#1}#2\end{array} \right ) }

\def\rd{\textrm{d}}

\renewcommand{\O}{\mathrm{O}}

\newcommand{\bes}{\begin{eqnarray}}
\newcommand{\ees}{\end{eqnarray}}

\def\dd{\mathrm{d}}

\def\nn{\nonumber}

\numberwithin{equation}{section}

\begin{document}

\title{\Large{\textbf{\sffamily Displacement versus velocity memory effects \\ from a gravitational plane wave}}}
\author{\sffamily Jibril Ben Achour\;$^{1,2,3}$ and Jean-Philippe Uzan\;$^{4,5}$}
\date{\small{\textit{$^{1}$ Arnold Sommerfeld Center for Theoretical Physics, Munich, Germany, \\
$^{2}$ Munich Center for Quantum Sciences and Technology, Germany,\\
$^{3}$ Univ de Lyon, ENS de Lyon, Laboratoire de Physique, CNRS UMR 5672, Lyon 69007, France,\\
 $^{4}$ Institut d'Astrophysique de Paris, UMR-7095 du CNRS, Paris, France, \\
 $^{5}$ Center for Gravitational Physics and Quantum Information,
Yukawa Institute for Theoretical Physics, Kyoto University, 606-8502 Kyoto, Japan. }}}

\maketitle

\begin{abstract}
This article demonstrates that additionally to the well-known velocity memory effect, a vacuum gravitational plane wave can also induce a displacement memory on a couple of test particles. A complete  classification of the conditions under which a velocity or a displacement memory effect occur is established. These conditions depend both the initial conditions of the relative motion and on the wave profile. The two cases where the wave admits a pulse or a step profile are treated. Our analytical expressions are then compared to numerical integrations to exhibit either a velocity or a displacement memory, in the case of these two families of profiles. Additionally to this classification,  the existence of a new symmetry of polarized vacuum gravitational plane wave under M\"{o}bius reparametrization of the null time is demonstrated. Finally, we discuss the resolution of the geodesic deviation equation by means of the underlying symmetries of vacuum gravitational plane wave.
\end{abstract}

\thispagestyle{empty}
\newpage
\setcounter{page}{1}

\hrule
\tableofcontents
%\addtocontents{toc}{\protect\setcounter{tocdepth}{2}} 
\vspace{0.7cm}
\hrule

%%%%%%%%%%%%%%%%%%%%%%%%%%%%%%%%%%%%%%%%%%%
\newpage

General relativity predicts that additionally to its oscillatory contribution, a gravitational wave contains also a non-oscillatory component which may  induce constant shifts in physical observables measured at early and late time after the passage of the wave. This effect was first noticed in the context of linearized gravity by Zel' dovich and Polnarev \cite{Zeldovich:1974gvh}, who showed that the relative distance between two test particles could be shifted after the passage of a GW. Their investigation was later extended to the full non-linear regime by Christodoulou \cite{Christodoulou:1991cr} and independently by Blanchet and Damour \cite{Blanchet:1992br}. Since then, a whole family of memory effects have been identified, from the velocity-kick memory \cite{Braginsky:1985vlg, Bieri:2024ios}, the spin memory \cite{Pasterski:2015tva} and the center-of-mass memory \cite{Nichols:2018qac}.  See also \cite{Seraj:2021rxd}.

Memory effects play a key role in our understanding of gravitational radiation. They stand as one of the last prediction of GR which has not yet been confirmed experimentally. They could nevertheless been observed in the coming years with the growing number of gravitational waves detections  or with the next generation of ground or space based interferometers so that the  prospects for their detection are  now attracting important efforts \cite{Lasky:2016knh, Boersma:2020gxx, Hubner:2021amk, Goncharov:2023woe}. From a theoretical perspective, memories effects are intimately related to the fine structure of the infrared regime of any gauge theory for which  the presence of soft radiative degrees of freedom induces a degeneracy of the vacuum. In the gravitational context, the presence of soft gravitons in an asymptotically flat region induces an infinite tower of memory effects which shift the null boundary observables, changing the physical infrared properties of the spacetime. The energy escaping through $\cI^{+}$ via gravitational radiation thus induces a transition from one vacuum to another which are related to each other by a soft diffeomorphism. This intimate relationship between gravitational memory, asymptotic symmetries and soft theorems, first recognized in  Ref.~\cite{Strominger:2014pwa} and formalized in the infrared triangle picture  \cite{Strominger:2017zoo}, opens a new avenue  to understand the properties of the infrared regime of gravity\footnote{See \cite{Hawking:2016msc, Donnay:2018ckb, Bhattacharjee:2020vfb, Sarkar:2021djs} for the role of soft diffeomorphisms and memories in the context of the black hole horizon.} \cite{Barnich:2009se, Campiglia:2014yka, Campiglia:2016efb, Compere:2019odm, Strominger:2021mtt, Freidel:2021fxf, Freidel:2021dfs, Blanchet:2023pce, Geiller:2024bgf}. Yet, some memory effects, dubbed persistent gravitational wave observables, do not fit in this elegant picture as they cannot be understood in terms of asymptotic symmetries of asymptotically flat spacetimes\footnote{The velocity-kick memory, first noticed long time ago by Braginsky and Grischuck \cite{Braginsky:1985vlg}, was related to the asymptotic super-boost symmetries for impulsive gravitational waves in \cite{Compere:2018ylh}.}\cite{Flanagan:2018yzh, Grant:2021hga, Seraj:2022qqj, Grant:2023ged, Siddhant:2024nft}.
Therefore, it appears useful to further investigate the realization of memories and their relationship to explicit and hidden symmetries beyond the context of asymptotically flat radiative spacetimes. 

An alternative arena in which  memory effects and their relation to symmetries can be studied are pp-wave geometries \cite{Roche:2022bcz}.  This family of exact radiative solutions of general relativity provides a simple enough model of non-linear vacuum gravitational wave  which allows one to analytically solve the motion of test particles and thus analyze memory effects. 
%At very large distance from the source, the front wave can be approximately described as a {\JPU plane and pp-wave provides thus a good approximation of the incident wave [passage pas clair. Reformuler]}. 
Moreover, as shown by Penrose, the leading tidal effects around a null geodesic in an arbitrary gravitational field can be described by a vacuum plane wave geometry through the so called Penrose limit \cite{PenroseLimit}. They also provide a simple framework to discuss boosted black holes and gravitational shock-waves \cite{Impulsive, Steinbauer:1997dw, Zhang:2017jma, Steinbauer:2018iis, Shore:2018kmt, Adamo:2022rob, He:2023qha}. Thus, pp-wave spacetimes represent a simplified but valuable model to analyze memory effects in the fully non-linear regime. So far, all investigations have confirmed that after the passage of a pp-wave, a couple of test particles with vanishing initial  relative velocity escape from each other at a constant relative velocity, providing a realization of the velocity memory effect\footnote{Spin memory in gyratonic gravitational plane wave have also been derived in Ref.~\cite{Shore:2018kmt}.} \cite{Zhang:2017rno, Zhang:2017geq, Zhang:2018srn, Zhang:2018gzn, Divakarla:2021xrd, Chakraborty:2022qvv, Elbistan:2023qbp,Chakraborty:2020uui}. However, it has been repeatedly argued that no displacement memory effect can take place in such vacuum pp-wave\footnote{See in particular the discussion in Section IV.E in Ref.~\cite{Zhang:2017geq}.} \cite{Zhang:2017rno, Zhang:2017geq}.  In view of the large freedom in choosing the wave profile while still solving the vacuum Einstein equation, it appears surprising that the displacement memory cannot be realized in a vacuum gravitational plane wave geometry. More recently, this conclusion was challenged  by Ref.~\cite{Zhang:2024uyp} who reported numerical examples of wave-profiles inducing a displacement memory. However, a clear understanding of the conditions under which this happens is missing. Thus, the first goal of this work is to provide a detailed analysis of the conditions under which a given wave-profile will induce a displacement or a velocity memory effect in the transverse direction. This is achieved both for the well-known case of pulse profiles and for step profiles which have not been studied so far.  As we shall see, deriving these general conditions allows us  to identify suitable examples where the displacement memory does occur. Appendix~\ref{Long} extends our analysis of transvervse memories to the conditions for having memories in the longitudinal direction.

Besides providing these general conditions to distinguish between velocity and displacement memories, it is also interesting to understand the interplay between these memories and the symmetries of pp-waves.  Memory effects are usually described by solving the geodesic deviation equation (GDE). In this framework, relating memories to the symmetries of the spacetime can be done by investigating the integrability of the GDE. As shown initially by Caviglia, Zordan, and Salmistraro (CZS) in Ref.~\cite{Caviglia0}, any rank-$n$ Killing tensor (and thus any Killing vector) can be used to construct an exact solution of the GDE; see also Refs.~\cite{Caviglia1, Salmistraro:1983xr, Dolan, Cariglia:2018erv} for an account on this result. In the case of pp-wave, the isometry group was derived long time ago in Refs.\cite{Souriau, Sippel:1986if, Maartens:1991mj, Keane:2004dpc}, where it was shown that any pp-wave possesses a five-dimensional group of isometries. It was later recognized that it corresponds to the Carrolian group in $2+1$  dimension without rotation \cite{Duval:2017els, Zhang:2019gdm}. Additionally to these isometries, a pp-wave also possesses a homothetic Killing vector which was shown to induce the existence of a non-trivial Killing tensor \cite{Keane:2010hg, Rani:2003br}. These symmetries allow one to completely integrate the geodesic motion in a generic pp-wave. In the following, we complete this well-known result by explicitly showing how the GDE can be solve by means of the different symmetry generators using the CZS theorem. This provides a way to label the relative motion between test particles by the conserved charges of the geodesic motion. 

Along the way, we shall also review, clarify and simplify several aspects in the analysis of vacuum gravitational plane waves. In particular, a well-known difficulty when investigating a vacuum gravitational plane wave is the description of the wave-profile in different coordinate systems. When switching from the Brinkmann to the Baldwin-Jeffrey-Rosen (BJR) coordinates, one ends up with a non-linear differential equation (\ref{wavBrink2}) relating the two wave-profiles even when one restricts to the $+$ or $\times$ polarization. We  revisit this complicated relation and provide two new results. First, we use a simple change of variable which allows one  to considerably simplify the non-linear relation which in turn simplifies the identification of the conditions for the different memories. Second, we demonstrate that the non-linear vacuum Einstein equations (written in BJR coordinates) enjoy a global symmetry under M\"{o}bius transformations of the wave profile, providing a solution-generating map for these simple radiative geometries.  This symmetry of non-linear vacuum gravitational plane wave provides a complementary result to the symmetry discussed recently in Ref.\cite{Zhao:2024xzo}.

This article is organized as follows. In Section-\ref{sec1}, we review the description of the vacuum gravitational plane wave geometry. We discuss how to efficiently describe the polarized case and present the M\"{o}bius invariance of the vacuum Einstein equations. In Section~\ref{sec2}, we review the derivation of the isometries in BJR coordinates, the existence of the HKV and the expression of the irreducible Killing tensor. Section~\ref{sec3} is devoted to the integration of the geodesic motion, a review of the construction of the adapted Fermi coordinates, following closely Ref.~\cite{Blau:2006ar}, and finally to the resolution of the parallel transport equation.  After having review this material, the different conditions leading to a displacement or a velocity memory effects, both in the transverse direction and in the longitudinal direction, are presented respectively  in Section~\ref{sec5} and Appendix~\ref{Long} together with explicit examples satisfying these conditions. Finally, Section~\ref{sec7} details the integration of the GDE by means of the symmetries, using the CZS theorem. To finish, Section~\ref{sec8} provides a discussion of our results and their implications.

\section{Generic properties of a gravitational plane wave}

In this section, we review the main properties of a gravitational plane wave and fix our notations. While most of the results presented in this first section are well-known, we clarify some of them and present also new formulations, exhibiting in particular a new symmetry of vacuum gravitational plane waves which does not seem to have been noticed so far.

Consider the metric of a gravitational plane wave in the so-called BJR coordinates,
\begin{align} \label{met}
\rd s^2 = 2 \rd u \rd v + A_{ij} (u)\rd x^i \rd x^j
\end{align}
with $A_{ij}$ a symmetric $2 \times 2$ matrix describing the wave profile such that 
\be\label{defAij}
A_{ij} =  \mat{cc}{A_{11} &  A_{12} \\ A_{12} & A_{22}}\,.
\ee
The Minkowski metric is trivially recovered by $A_{ij} = \delta_{ij}$.  This radiative geometry is defined by exhibiting a covariant constant null vector $\xi^{\alpha} \partial_{\alpha} = \partial_v$, i.e. which satisfies $\nabla_{\alpha} \xi_{\beta} =0$. The BJR coordinates are usually not those used to investigate the properties of the wave as they are not globally defined and thus fail to cover the whole spacetime, as discussed in Ref.~\cite{Wang:2018iig}. Within this coordinate system, the Einstein field equations take the form
\begin{align}\label{e.EinstEq}
 \partial_u \left( A^{i\ell} \partial_u A_{i\ell}  \right) + \frac{1}{2}  A^{i\ell}  A^{j k}\partial_u A_{j\ell}   \partial_u A_{i k}  = 8\pi T_{uu}
\end{align}
where $T_{uu}$ is stress-energy tensor of some source.  As it is well-known, this field equation takes a much simpler form in the Brinkmann coordinates system which we shall discuss later.   However, in the vacuum and in the case of polarized wave for which $A_{12} =0$, this non-linear equation can be dramatically simplified; see Appendix~\ref{appB} for a discussion on the polarisations. Indeed, in such a case, $A^{ii} = 1/A_{ii}$ and field equation reduces to
\begin{align}
\label{fieldeq}
 \left[ \frac{\ddot{A}_{11}}{A_{11}}  - \frac{1}{2} \left(\frac{\dot{A}_{11}}{A_{11}} \right)^2 \right]  +\left[ \frac{\ddot{A}_{22}}{A_{22}}  - \frac{1}{2} \left( \frac{\dot{A}_{22}}{A_{22}} \right)^2\right]=0\,.
\end{align}
In general, solving this non-linear equation is challenging so that most of the analysis rely on numerical investigations. However, it can be dramatically simplified by setting
\be
A_{ii}(u) \equiv {\cal A}^2_i(u)\,,
\ee
which allows one to write
$$
 \frac{\ddot{A}_{ii}}{A_{ii}} - \frac{1}{2} \left( \frac{\dot{A}_{ii}}{A_{ii}}\right)^2 = \frac{2\ddot {\cal A}_i}{{\cal A}_i}
$$
while the dynamical equation (\ref{fieldeq}) simplifies to
$$
\frac{\ddot {\cal A}_1}{{\cal A}_1} + \frac{\ddot {\cal A}_2}{{\cal A}_2} =0\,.
$$
With this notation, the polarized wave-profile takes the form
$$
A_{ij} =  \mat{cc}{{\cal A}_1^2&  0 \\ 0 & {\cal A}^2_2 } \,.
$$
As we shall see, working with $\A_i$ instead of $A_{ii}$ allows us to write the conditions for the different memory effects in a very compact form. Before presenting these results, let us show that the non-linear wave equation (\ref{fieldeq}) can be reformulated so as to reveal a hidden symmetry.

\subsection{Conformal invariance of the wave profile}

\label{sec1}

In order to make this hidden symmetry explicit, it is useful to introduce the auxiliary fields 
\be
A_{11} = \frac{1}{\dot{F}} \qquad A_{22} =\frac{1}{\dot{G}}\,.
\ee
Then, one can recast the field equations as
\begin{align}
\label{sch}
 \text{Sch}[F] +  \text{Sch}[G] =0 
\end{align}
where $ \text{Sch}[f] $ denotes the Schwarzian derivative of the function $f$ defined by 
\be
 \text{Sch}[f] = \frac{\dddot{f}}{\dot{f}} - \frac{3}{2} \frac{\ddot{f}^2}{\dot{f}^2} \,.
\ee
The Schwarzian derivative being M\"{o}bius invariant, i.e. for any given function $f$, 
\be
\text{Sch}[M\circ f] =  \text{Sch}[f] \qquad \text{where} \qquad M (u) = \frac{a u + b}{c u + d} \qquad ad-bc\neq 0\,,
\ee
it follows that knowing one couple of solution $(F,G)$, one can generate another inequivalent solution to the wave equation by a M\"{o}bius transformation of the seed profile\footnote{It is also useful to rewrite the Schwarzian derivative as follows
\be
 \text{Sch}[f] = \frac{\rd^2}{\rd u^2} \log{f'} - \frac{1}{2} \left(\frac{\rd}{\rd u} \log{f'}\right)^2\,.
\ee where $f' := \frac{\rd f}{\rd u}$}. In general, the Einstein equations can be understood as a balance equation between the Schwarzian derivative of each component of the polarized wave-matrix $A_{ij}$. 

The simplest wave profile solving the field equation (\ref{sch}) corresponds to the case where each Schwarzian derivative vanishes independantly, which is solved for
\be
\label{FG}
F (u) = \frac{a u +b}{c u + d} \,,\qquad  G(u) = \frac{\tilde{a} u +\tilde{b}}{\tilde{c} u + \tilde{d}}
\ee
where the eight constants satisfy  $ad -bc \neq 0$ and $\tilde{a} \tilde{d} -\tilde{b} \tilde{c} \neq 0$
such that
\be
\label{solquad}
A_{11} (u) = (c u + d)^2\,,  \qquad A_{22} (u) = (\tilde{c} u + \tilde{d})^2\,.
\ee
where the constants in Eq~(\ref{solquad}) are rescaled versions of the constants entering in Eq~(\ref{FG}).
This specific solution corresponds to one of the simplest example of a vacuum gravitational plane wave. Notice that while the profile is non-trivial, such a plane wave geometry can always be brought back to the Minkowski  metric (in light-cone coordinate) thanks to the following diffeomorphism
\begin{align}
& u \rightarrow u \,, \\
 & v \rightarrow  v - \frac{1}{2} c (cu+d) x^2 - \frac{1}{2} \tilde{c} (\tilde{c}u+\tilde{d}) y^2\,, \\
 &  x \rightarrow (cu +d) x\,, \\
 &  y \rightarrow (\tilde{c}u+\tilde{d}) y\,.
\end{align}
This diffeomorphism can be considered as a large gauge transformation.
From the point of view of the solution generating map, the simple solution (\ref{solquad}) stands as a fixed point of the solution space, as it remains invariant under a M\"{o}bius transformation. 

Beyond this simple example, an infinite set of solutions for the wave profile exists, which are simply constrained by the balance equation (\ref{sch}) between the two Schwarzian derivatives. In the following, we shall be interested in distinguishing between the wave profile that induces a constant displacement memory and those that instead induce a constant velocity memory. Having presented the structure of the vacuum field equations, we now turn to the symmetries of the vacuum plane wave metric.

\subsection{Symmetries}\label{sec2}

The symmetries of pp-wave have already been investigated and classified long time ago \cite{Sippel:1986if, Maartens:1991mj, Keane:2004dpc, Duval:2017els}. It was shown by Souriau~\cite{Souriau} that one can express the conformal Killing vector (KV) fields in a close form provided one uses the BJR coordinates;  see Ref.~\cite{Zhang:2019gdm} for a recent review of this result. Consider the metric (\ref{met}) \textit{without} the restriction $A_{12} =0$. Consider then the conformal Killing equation
\be
\cL_{\xi} g_{\mu\nu} = \Omega g_{\mu\nu}\,,
\ee
where $\Omega(u,v,x^i)$ is the conformal factor. In BJR coordinates, it decomposes as
\begin{align}
\partial_v \xi^{u} & =0\,,   \label{e.kv1} \\
\partial_u \xi^v & =0 \,,  \\
\partial_i \xi^u + A_{ij} \partial_v \xi^j & =0\,,  \\
 \partial_i \xi^v + A_{ij} \partial_u \xi^j  &= 0\,,  \\
\partial_v\xi^v + \partial_u \xi^u  &= \Omega \,,    \\
 \xi^u \partial_u A_{ij} + A_{ik} \partial_j \xi^k + A_{jk} \partial_i \xi^k   &= \Omega A_{ij} \label{e.kv6} \,.  
\end{align}
It is straightforward to check that the vector fields
\be\label{e.defT}
\cN^{\alpha} \partial_{\alpha} = \partial_v \;, \qquad P_{+}^{\alpha}\partial_{\alpha} = \partial_x \qquad P^{\alpha}_{-} \partial_{\alpha} = \partial_y
\ee
solve these Killing equations~(\ref{e.kv1}-\ref{e.kv6}). Moreover, using the Souriau matrix defined by
\be
H^{ij} (u_0,u) \equiv \int^u_{u_0} A^{ij} (w) \rd w
\ee
one can show that the two additional vector fields given by
\begin{align}\label{BC}
\B^{\alpha}_{+} \partial_{\alpha} & = H^{xx} (u_0,u)\partial_x + H^{xy} (u_0,u)\partial_y - x  \partial_v \\
\B^{\alpha}_{-} \partial_{\alpha} & = H^{yy}(u_0,u) \partial_y + H^{yx} (u_0,u)\partial_x -  y \partial_v 
\end{align}
also solve the Killing equations~(\ref{e.kv1}-\ref{e.kv6}). This completes the number of Killing vector fields for this spacetime geometry. They can be written in a compact form as
\begin{align}
\xi^{\alpha} \partial_{\alpha} = \left\lbrace h + b_i x^i  \right\rbrace \partial_v + \left[\chi^i + b_j H^{ij} (u_0,u)\right] \partial_i \,.
\end{align} 
The isometry group is thus parametrized by the five constants $(h, \chi^i, b_i)$. As we shall see in the last section, these gauge parameters can be directly linked to the initial conditions of the geodesic deviation and shown to label the different contributions to the memories. 

Let us emphasize that we never had to explicitly fix the wave profile $A_{ij}$ to obtain the Killing vectors in a closed form. This implies that these Killing vector fields are a common properties of \textit{any gravitational plane wave} described by the metric~(\ref{met}).  We stress that identifying the full set of symmetries of the plane gravitational wave is greatly simplified by working in BJR coordinates~(\ref{met}), where the metric appears to be homogeneous in the $u$-coordinate.  Computing the Lie algebra of these vector fields show that the only non-vanishing bracket is
\be
[B_{\pm}, P_{\pm} ] = \cN\,,
\ee
which can be identified with the Carrolian Lie algebra in $2+1$ dimensions without rotation.

Let us now consider the case $\Omega \neq 0$. Indeed, it is well-known that any pp-wave possesses a homothetic Killing vector (HKV) rescaling the wavefront. One can show that for general wave profile $A_{ij}$, the only solution to the conformal Killing equation is $\Omega =2$ while the conformal vector field takes the form
\begin{align}\label{e.HKV}
Z^{\alpha} \partial_{\alpha} & = 2v \partial_v +  x^i \partial_i \,.
\end{align}
Computing its brackets with the KVs, one finds that it acts as a dilatation through
\begin{align}
[P_{\pm}, Z ]  = P_{\pm}\,, \qquad  [B_{\pm}, Z ]  =  B_{\pm}\,, \qquad [\cN, Z ] & = 2\cN \,.
\end{align}
Quite remarkably, the existence of this HKV allows us to further identify an additional hidden symmetry.

Indeed, as initially shown by Koutras \cite{Koutras}, if a spacetime admits both a gradient Killing vector and a HKV then it also admits a non-trivial rank-2 Killing tensor (KT) which generates an additional symmetry \cite{Rani:2003br}. With $\xi_{\alpha} \rd x^{\alpha}=(\partial_\alpha\Phi)\dd x^\alpha$ the gradient KV  and $Z_{\alpha} \rd x^{\alpha}$ the HKV, the KT is explicitely given by
\be
K_{\mu\nu} \rd x^{\mu} \rd x^{\nu} = \left[  Z_{(\mu} \xi_{\nu)} -  \Phi g_{\mu\nu} \right] \rd x^{\mu} \rd x^{\nu}.
\ee
In the case we consider, in which the gradient KV is the covarianty constant null vector, ii.e. $\cN^{\alpha} \partial_{\alpha} =\partial_v$, the KT is given by
\begin{align} \label{KTBJRR}
K_{\mu\nu} \rd x^{\mu} \rd x^{\nu} & =  2v \rd u^2  - u ( 2 \rd u \rd v + A_{ij} \rd x^i \rd x^j) + A_{ij} x^j  \rd u \rd x^i \,.
\end{align}
Note that its expression holds, again, for any wave profile. It corresponds to the Killing tensor described by Keane and Tupper in Ref.~\cite{Keane:2010hg}. As we shall see, the fact that the Killing tensor is not a symmetrized product of Killing vectors implies that it provides a new charge for the geodesic motion. We will now use these symmetries to analytically solve both the geodesic motion and the geodesic deviation. This will reveal useful when investigating the memory effects.

\subsection{Geodesic motion}

\label{sec3}

Let us now analyze the geodesic motion using BJR coordinates. Consider a geodesic $x^\mu(\tau)$ with affine parameter $\tau$ and tangent vector $u^\mu=\rd x^\mu/\rd\tau$ normalized to 
\begin{equation}\label{e.dfeps}
u^\mu u_\mu=\epsilon\quad \hbox{with}\quad \epsilon = \{ 0,-1\}
\end{equation}
respectively for null and timelike geodesics.  For the metric~(\ref{met}), the geodesic Lagrangian, $L = \frac{1}{2} g_{\mu\nu} \dot{x}^{\mu} \dot{x}^{\nu}$,  reduces to
\begin{align}
L= \dot{u}\dot{v}  + \frac{1}{2} A_{ij} \dot{x}^i \dot{x}^j
\end{align}
with a dot referring to a derivative w.r.t. to the proper time $\tau$. The geodesic equation $u^{\mu} \nabla_{\mu} u^{\nu} = 0$ decomposes into
\begin{align}\label{eom2}
\ddot{v}   - \frac{1}{2}  A'_{ij} \dot{x}^i \dot{x}^j & =0 \,, \\
\label{eom1}
A_{ij} \ddot{x}^j +A'_{ij} \dot{x}^j  \dot{u}   & =0 \,,\\
\label{eom3}
\ddot{u} & =0\,,
\end{align}
with a prime refering to a derivative w.r.t. $u$. In the following, we review how to algebraically integrate these equations using the conserved charges associated to the underlying symmetries, including the KT charge.

\subsubsection{Conserved charges}

Let us first write the Hamiltonian formulation of this system in order to discuss its symmetries. The momenta are given by
\begin{align}
\label{e.139}
p_u &= \frac{\delta \cL}{\delta \dot{u}} =  \dot{v}  \,, \\
\label{e.140}
p_i  &= \frac{\delta \cL}{\delta \dot{x}^i} =  A_{ij} \dot{x}^j  \,, \\
\label{e.141}
 p_v &= \frac{\delta \cL}{\delta \dot{v}} = \dot{u}  \,,
\end{align}
such that the symplectic structure reads
\be
\{ v, p_v\} =\{ u, p_u\} =1\,, \qquad \{ x^i, p_j\} = \delta^i{}_j\,.
\ee
Then, the Hamiltonian of the geodesic motion takes the form
\begin{align}
\label{e.143} H& = p_u p_v  +\frac{1}{2} A^{ij} p_i p_j = \frac{\epsilon}{2}\,.
\end{align}
This system admits six conserved charges associated to the symmetries identified in Section~\ref{sec2}. Let us recall that for each isometry generated by $\xi^{\alpha} \partial_{\alpha}$, the scalar $\cO = \xi^{\alpha} p_{\alpha}$ provides the associated conserved charge. Moreover, for the Killing tensor $K_{\mu\nu}$ the charge is quadratic in the momenta and reads $K = K^{\mu\nu} p_{\mu} p_{\nu}$. In our case, one finds the following conserved quantities:
\begin{itemize}
\item {\it Translations}. Since the Hamiltonian does not depend on neither $v$ nor $x^i$, $p_v$ and $p_i$ are automatically conserved. We denote them as
\be
\cN = p_v\,, \qquad P_{+} = p_x \,, \qquad P_{-} = p_y\,,
\ee
which provide 3 constants of motion associated to the KV~(\ref{e.defT}).
\item {\it Carrolian boost}. The conserved charges generating the boosts~(\ref{BC}) are given by
\begin{align}
\label{e.BP}\B_{+} & = H^{xx} p_x + H^{xy} p_y - p_v x \,, \\
\label{e.BM}\B_{-} & = H^{yy} p_y + H^{yx} p_x - p_v y\,.
\end{align}
A simple computation shows that
\begin{align}
\{ \B_{\pm}, H \} &=0 \,,
\end{align}
so that these new charges are indeed conserved.  These two charges allow one to integrate the geodesic motion.
\item  {\it Conformal charge}. The HKV~(\ref{e.HKV}) is associated to the charge
\begin{align}\label{e.148}
\cZ  & =  p_v 2v  + p_i x^i\,,
\end{align}
the bracket of which with the Hamiltonian reduces to
\begin{align}
\{ \cZ, H\} & = 2 H\,.
\end{align}
This charge is conserved only when $H=0$, i.e. for null geodesics, as expected for a conformal Killing vector. 
\item  {\it Hidden Killing tensor charge}. The charge coming from the Killing tensor~(\ref{KTBJRR}) can be expressed in a simple form as
\begin{align} \label{KTC}
\cK = K^{\mu\nu} p_{\mu} p_{\nu} & =  p_v \cZ - 2 u H \,.
\end{align}
As expected, it is independent of the charges inherited by the conformal isometries.
\end{itemize}

From the explicit expressions of the six charges $(\cN, P_{\pm}, \B_{\pm},\cK)$, we can now compute their algebra.  First, one finds that the five charges $(\cN, P_{\pm}, \B_{\pm})$ form a $2+1$ Carrolian algebra without rotation \cite{Zhang:2019gdm}, i.e the only non-vanishing bracket reads
\begin{align}
\{ P_{\pm}, \B_{\pm}\} = \cN\,.
\end{align}
This is the well-known 5-parameters isometry group of the gravitational plane wave. Then, the Killing tensor charge $\cK$ builds up the conformal extension of the Carrolian algebra given by
\begin{align}
\{ P_{\pm}, \cK \}  = \cN P_{\pm}\,, \qquad  \{ B_{\pm}, \cK \}  =  \cN B_{\pm}\,, \qquad \{\cN, \cK \} & = 2\cN^2\,,
\end{align}
which closes the algebra of conserved charges. The system is integrable in the sense that the four charges $(\cN, \B_{\pm}, H)$ are in involution.

\subsubsection{Algebraic integration of geodesic flow}

Now, let us review how these conserved charges allow for an algebraic integration of the geodesic equation~(\ref{eom1}-\ref{eom3}).  First, we use the conservation of $\cN$, i.e. of $p_v$, to integrate the $u$-trajectory equation~(\ref{eom3}) or equivalently~(\ref{e.141}) and write 
\begin{equation}\label{utraj}
u = p_v \tau = \cN \tau.
\end{equation} 
Then, using the two charges $\B_\pm$ generating the Carrolian boosts, the expressions~(\ref{e.BP}-\ref{e.BM}) directly give the $x$  and $y$ trajectories as
 \begin{align}
 x(u) & = \frac{1}{\cN} \left[  H^{xx} (u_0,u) P_{+} + H^{xy} (u_0,u) P_{-}\right]  - \frac{\B_{+}}{\cN}\,, \\
 y(u) & = \frac{1}{\cN} \left[  H^{yy} (u_0,u) P_{-} + H^{yx} (u_0,u) P_{+} \right]   - \frac{\B_{-}}{\cN} \,,
\end{align}
which combine to give the general solution of Eq.~(\ref{e.140}) as
\begin{align}
\label{xtraj}
x^i (u) & = \frac{1}{\cN}\left[ H^{ij}(u_0,u) p_j - \B^i \right] ,
\end{align}
with $\B^i = (\B_{+}, \B_{-})$. This provides the link between these two conserved charges and the initial conditions as
\begin{align}
\label{e.bxO}
x^i_0\equiv x^i (u_0) & = -\frac{\B^i}{\cN}\,.
\end{align}
To finish, the conservation of the charge $\cK$~(\ref{KTC}) associated to the KT gives
\be
p_v\cZ = \epsilon u + \cK 
\ee
once Eq.~(\ref{e.143}) is used to express $H$ in terms of $\epsilon$. From the expression~(\ref{e.148}) of $\cZ$, once $x^i$ is extracted from Eq.~(\ref{xtraj}) and, keeping in mind that $p_v=\cN$, one algebraically gets the $v$-trajectory as
\begin{align}\label{vtraj}
v(u) & =  \frac{1}{2\cN^2} \left[ \epsilon u  -  H^{ij}(u_0,u) p_i p_j    + p_i \B^i + \cK \right].
\end{align}
This provides the link between $\cK$ and the initial conditions as
\begin{align} \label{e.bvO}
v_0\equiv v(u_0) & = \frac{1}{2\cN^2}\left[ \epsilon u_0  -\cN p_i x^i_0 + \cK \right].
\end{align}
The set of equations~(\ref{utraj}, \ref{xtraj}, \ref{vtraj}) provides an algebraic solution of the geodesic equation in terms of the 5 conserved charges $(\cN, \B^i, \cK)$ once the integrals $H^{ij}(u_0,u)$ are determined. As seen from Eqs.~(\ref{e.bxO}) and~(\ref{e.bvO}), the 3 constants  $(\B^i, \cK)$ are related to the initial conditions $(x^i_0, v_0, p_i)$ while Eq.~(\ref{utraj}) shows that $\cN$ can be chosen arbitrarily. Notice that since $\epsilon$ appears only in Eq.~(\ref{vtraj}), the difference between the time-like and null geodesics only shows up in the longitudinal motion, i.e. in the $v$-trajectory.\\

This solution allows us to derive the expression of the 4-velocity $u^{\mu} \partial_{\mu}$. In full generality, its components are given by
\begin{align}
u^u & = \frac{\rd u}{\rd \tau} = \cN   \nn \,,\\
u^i & =  \frac{\rd x^i}{\rd \tau} = A^{ij} p_j  \nn \,, \\
u^v & =  \frac{\rd v}{\rd \tau} = \frac{1}{2 \cN} \left(\epsilon    - A^{ij} p_ip_j \right)\,.
\end{align}
One can check that it satisfies, as expected, $u^{\mu} u_{\mu} = \epsilon$ and $u^{\mu} \nabla_{\mu} u^{\nu} =0$. With this solution, one can investigate the properties of any general congruence of timelike geodesics. 

To finish, this expression allows us to derive the optical scalars associated to the congruence of time-like geodesics. Using that the velocity vector $u_{\mu}$ naturally defines a projector $h_{\mu\nu} = g_{\mu\nu} + u_{\mu} u_{\nu}$ such that $u^{\mu}h_{\mu\nu} =0$, one can then define the expansion, shear and twist of the congruence respectively as
\begin{align}
\Theta\equiv \nabla_{\mu} u^{\mu}\qquad \sigma_{\mu\nu} = \nabla_{(\mu} u_{\nu)}- \frac{\Theta}{3} h_{\mu\nu} \qquad \omega_{\mu\nu} = \nabla_{[\alpha} u_{\beta]} 
\end{align}
In the case of a congruence of time-like geodesics in our gravitational plane wave, one obtains
\begin{align}
\Theta &= \cN  \varrho \,  \\
\sigma & = - \cN^2 \left[ \dot{A}^{ij} \dot{A}_{ij} + \frac{2}{3}\varrho (\varrho + \dot{A}_{ij} p^i p^j ) - \frac{1}{9} \varrho^2 (p_i p^i)^2\right]  \,,
\end{align}
where where $\sigma = \sigma_{\mu\nu} \sigma^{\mu\nu}$ is the squared invariant shear and where we have introduced the quantity $\varrho = A^{ij} \dot{A}_{ij}$.
The rotation (or vorticity) tensor identically vanishes, i.e. $\omega_{\mu\nu}  =0$, showing that the tangent geodesic vector $u^{\alpha} \partial_{\alpha}$ is hyper-surface orthogonal. The only effects of the wave reduce to an isotropic expansion/contraction and shearing of the bundle of time-like geodesics. This concludes our analysis of the geodesic motion. 

\subsection{Adapted Fermi coordinates }

In order to discuss the memory effects, one first need to construct a set of adapted Fermi normal coordinates (FNC). We refer the interested reader to Refs \cite{Manasse:1963zz, Marzlin:1994ia, Marzlin:1994wc, Delva:2011abw, Guedens:2012sz} for details on the general procedure and to Refs \cite{Baskaran:2003bx, Rakhmanov:2004eh, Delva:2006qa, Rakhmanov:2008is, Rakhmanov:2009zz, Rakhmanov:2014noa, Blau:2006ar} for its application to gravitational wave physics. To proceed,  one needs to introduce suitable tetrad fields. Both the construction and the evolution of this tetrad are discussed in this section. Without loss of generality, we set 
$$
\cN=p_v=1\,
$$ 
which is only a choice of the units of the $u$ coordinate. This implies that $u=\tau$ so that a dot shall now refer to a derivative with respect to either $\tau$ or $u$.

The starting point is to consider a reference geodesic $\bar{\gamma}$ characterized by a vanishing translational charge, i.e. $p_i =0$. Its tangent vector and dual co-vector are given by
\begin{align}
\label{velo}
\bar{u}^{\mu} \partial_{\mu}  = \partial_u \qquad \bar{u}_{\mu} \rd x^{\mu}  = \rd v  \,.
\end{align}
Here, we focus on a null reference geodesic. The reason is that we shall analyze the memory effects in the transverse plane for which the motion does not depend on the value of $\epsilon$. This parameter only affects the longitudinal motion. The equation of the  null-geodesic $\bar{\gamma}(u)$ with affine parameter $u$ and initial conditions $(\bar{v}_0,\bar{x}^i_0)$ in BJR coordinates derives trivially from Eqs.~(\ref{xtraj}) and~(\ref{vtraj}) (since $\epsilon=0$)
\begin{align}
\label{trajrefgeneric}
\bar{v}(u) & =  v_0 , \qquad \bar{x}^i(u) = \bar{x}^i_0  \,.
\end{align}
The geodesic with $(v_0,\bar{x}^i_0)=(0,0)$  corresponds, making use of  Eqs.~(\ref{e.bxO}) and~(\ref{e.bvO}), to the values of the conserved charges given by
\be
\label{chargeref}
\bar{v}_0 = \frac{1}{2}\bar{\cK} =0 \,,  \qquad \bar{\B}^i = -\bar{x}^i_0 =0 \qquad  \bar{p}_i =0\,
\ee 
so that
\begin{align}
\label{trajref}
\bar{v}(u) & = 0\,, \qquad \bar{x}^i(u) =  0 \,.
\end{align}
Now, let us introduce a set of adapted Fermi coordinates, $X^{I}$, with $I\in \{ 0,\ldots,3\}$. They are related to the BJR coordinates, $x^{\mu}$, by the tetrad $E^I{}_{\mu}$ defined by
\be
E^I{}_{\mu} \equiv \frac{\partial X^I}{\partial x^{\mu}} \,.
\ee
We construct this tetrad by imposing that on the geodesic of reference, the coordinate $X^0$ coincides with the affine parameter of the geodesic, i.e.
\be
\bar{E}^0{}_{\mu} \rd x^{\mu}= \bar{u}_{\mu}\rd x^{\mu} \,.
\ee
Then, we impose that the remaining components of the tetrad be parallel transported along the geodesic $\bar{\gamma}$, i.e. $\bar{u}^{\mu} \nabla_{\mu} E^I{}_{\nu} =0$, and that they satisfy the following orthogonality relations
\be
\label{orto}
g_{\mu\nu} |_{\bar{\gamma}} = \bar E^I{}_{\mu} \bar E^J{}_{\nu} \eta_{IJ} \qquad \text{with} \qquad \eta_{IJ} = \mat{cccc}{0&1&0&0\\1&0&0&0 \\ 0&0& 1 &0 \\ 0&0&0&1}\,.
\ee
This orthogonality conditions impose that
\begin{align}
\bar{E}^0{}_{\mu} \rd x^{\mu} & = 2\rd v  \,, \qquad \qquad \qquad \qquad\qquad   \;\; \bar{E}^{\mu}{}_{0} \partial_{\mu}  = \partial_u  \,,\\
\bar E^1{}_{\mu} \rd x^{\mu} & =  \rd u  \,,\qquad \qquad  \qquad \qquad \qquad \;\;\; \;\bar{E}^{\mu}{}_{1} \partial_{\mu}  =  \partial_v \,,\\
\bar E^A{}_{\mu} \rd x^{\mu} & = E^A{}_i \rd x^i  \,,\qquad \qquad \qquad \qquad  \;\;\;  \bar{E}^{\mu}{}_{A} \partial_{\mu}  = E^i{}_A \partial_i \,.
\end{align}
Since all the quantities depend on $u$ along, we have $E^I_\mu=\bar E^I_\mu$. Then, the parallel transport $\bar{u}^{\mu} \nabla_{\mu} E^I{}_{\nu} =0$ translates to
\be
\label{PT}
\dot{E}^A{}_{i} = \frac{1}{2} A^{jk} \dot{A}_{ki} E^A_j \equiv {S^j}_i {E}^A{}_{j} \,,
  \qquad 
 \dot{E}^{i}{}_A = - \frac{1}{2} A^{ik} \dot{A}_{kj} E^j{}_A = -{S^i}_j E^j{}_A \,.
\ee
Following the procedure described in Ref.~\cite{Blau:2006ar}, the null Fermi coordinates can be constructed by expanding their expressions in the space transverse to the reference geodesic. Let us denote  the Fermi coordinates by $X^0 =U$, $X^1 =V$ and $X^A = \{ X,Y\}$ such that the transverse space w.r.t. the reference geodesic admits the coordinates $X^a \in \{ V, X^A\}$. Then,  the Taylor expansion of the BJR coordinates up to second order reads
\begin{align}
x^{\mu}(U,V, X^A) = x^{\mu}(U) + \bar{E}^{\mu}{}_{a} (U) X^a - \frac{1}{2}  \bar{E}^{\alpha}{}_{a} (U)  \bar{E}^{\beta}{}_{b} (U) \Gamma^{\mu}{}_{\alpha\beta}(U)  X^a X^b\,.
\end{align}
It follows that the Fermi and BJR coordinates are related by
\begin{align}
u & = U \,,\\
v & =  V + \frac{1}{4} \dot{A}_{ij} \bar{E}^i{}_A \bar{E}^j{}_B X^A X^B \,, \\
\label{xFNC}
x^i & = \bar{E}^i{}_A X^A\,.
\end{align}
By construction, setting $V=0$ and $X^A =0$, one recovers the position (\ref{trajref}) of the reference geodesic $\bar{\gamma}$.
Inverting these relations, one obtains the Fermi coordinates 
\begin{align}
U & = u \, \\
\label{FNC2}
V & = v - \frac{1}{4} \dot{A}_{ij} x^ix^j  \,,\\
\label{FNC3}
X^A & = \bar{E}^A{}_i x^i \,.
\end{align}
One can easily check that the conditions (\ref{trajref}) recover the reference geodesic, i.e. $V=0$ and $X^A=0$, as it should. The metric of the gravitational plane wave in these Fermi coordinates reads
\begin{align}
\rd s^2 = 2 \rd U \rd V + \delta_{AB} \rd X^A \rd X^B + H_{AB}(U) X^A X^B \rd U^2\,,
\end{align}
with
\be
\label{FERMIPROF}
H_{AB} = \frac{1}{2}  E^i{}_{ A} \; \partial_u \left( \dot{A}_{ij} E^j{}_B\right) \,.
\ee
Within these Fermi coordinates, the vacuum field equation (\ref{fieldeq}) dramatically simplifies and reduces to the condition that $H_{AB}$ remains traceless, i.e. $\delta^{AB} H_{AB} =0$. Hence, one can pick up any wave profile of the form
\be
H_{AB} = \mat{cc}{H_{+} & H_{\times} \\ H_{\times} & - H_{+}}
\ee 
without any further restriction, providing a large freedom on the wave profile one can consider. Notice that this freedom in the choice of wave profile is much less transparent in the BJR coordinates we started with. 

At this level, the last step to complete the construction of the Fermi coordinates requires to solve the parallel transport condition. 

\subsection{Solving the transport parallel condition}

Consider the transport parallel (PT) equation
\be
\label{PT}
\dot{E}^{i}{}_A = - \frac{1}{2} A^{ik} \dot{A}_{kj} E^j{}_A\,.
\ee
In order to provide an analytic expression of the tetrad, we focus our analysis on polarized plane wave for which $A_{12} =0$ so that
\begin{align}
\label{prof}
A_{ij} =  \mat{cc}{A_{11} & 0 \\ 0 & A_{22}} \,.
\end{align}
Hence, the PT equation decomposes into
\begin{align}
 \dot{E}^1{}_A  = - \frac{1}{2} \frac{\dot{A}_{11}}{A_{11}} E^1{}_A\,, \qquad \dot{E}^2{}_A= - \frac{1}{2} \frac{\dot{A}_{22}}{A_{22}} E^2{}_A \,.
\end{align}
Using the redefinition $A_{ii} = \A_{i}^2$, these two equations integrate as
\begin{align}
E^1{}_A (u) 
= \frac{\A_{1}(u_0)}{\A_{1}(u)}E^1{}_A (u_0) \,, \qquad 
E^2{}_A (u) = \frac{\A_{2}(u_0)}{\A_{2}(u)}E^2{}_A (u_0)\,.
\end{align}
In order to write down the relation (\ref{FERMIPROF}) between the wave profile $H_{AB}$ in Binkmann coordinates and the wave profile $A_{ij}$ in BJR coordinates, it is useful to compute the dynamics of the tetrad. One obtains
\begin{align}
\dot{E}^1{}_A = - \frac{\dot{\A}_1(u)}{\A^2_1(u)} \A_1(u_0) E^1{}_A(u_0)\,,
\qquad \dot{E}^2{}_A  =  - \frac{\dot{\A}_2(u)}{\A^2_2(u)} \A_2(u_0) E^2{}_A(u_0)
\end{align}
and
\begin{align}
\ddot{E}^1{}_A 
= - \left( \ddot{\A}_1 - \frac{2 \dot{\A^2_1}}{\A_1}\right)\frac{ \A_1(u_0)}{\A^2_1(u)} E^1{}_A(u_0)\,,\qquad
 \ddot{E}^2{}_A =  - \left( \ddot{\A}_2 - \frac{2 \dot{\A^2_2}}{\A_2}\right)\frac{ \A_2(u_0)}{\A^2_2(u)} E^2{}_A(u_0)\,.
\end{align}
Now, using Eq.~(\ref{FERMIPROF}), one finds that $H_{\times} =0$, as expected, while $H_{+}$ is given by 
\begin{align}
\label{wavBrink1}
H_{+} =  \frac{\ddot{\A}_1}{\A_1}  = - \frac{\ddot{\A}_2}{\A_2} \,.
\end{align}
For completeness, we also give this relation in term of the BJR wave profile $A_{ii}(u)$,
\begin{align}
\label{wavBrink2}
H_{+} = \frac{1}{2} \left[  \frac{\ddot{A}_{11}}{A_{11}} - \frac{1}{2} \left( \frac{\dot{A}_{11}}{A_{11}} \right)^2\right] = -\frac{1}{2} \left[ \frac{\ddot{A}_{22}}{A_{22}} - \frac{1}{2} \left( \frac{\dot{A}_{22}}{A_{22}} \right)^2 \right]\,.
\end{align}
This provides the direct relation between the wave profile $H_{+}$ in Brinkmann and the wave profile $A_{ij}$ (or $\A_i$) in the BJR coordinates in the polarized case. Noticed that since $H_{+}$ can be chosen freely, the relation (\ref{wavBrink2}) provides a rather complicated non-linear equation one has to solve to determine the associated wave profile $A_{ij}$ in BJR coordinates. In general, picking up a profile $H_{+}$ and determining the associated profile $A_{ij}$ can only be done numerically\footnote{However, one can also first choose the function $A_{11}$ and determine $H_{+}$. Then, one has to solve the remaining non-linear equation to obtain $A_{22}$. While the this second strategy does not remove the difficulty to find $A_{22}$, one can choose the initial conditions for the geodesic motion such that the function $A_{22}$ does not play any role. Thus, this second approach allows one to keep control on the relevant functions when studying the geodesic motion and the associated memory effects.}. A more direct approach (when working in the polarized case as done here) is instead to choose $H_{+}$, then determine $\A_i$ by integrating the much simpler relation (\ref{wavBrink1}) and then determine $A_{ij}$. 

Before analyzing the conditions for the different memory effects, let us stress that one can also consider the second polarization where only $H_{\times} \neq 0$. As shown in appendix~\ref{appB}, this latter case can be brought back to the case studied in this section by a suitable constant rotation of the parallel transported tetrad.   For this reason, we focus on the case $H_{+}\neq 0$ only which is sufficient to analyze the different memory effects. 

\section{Transverse memories: displacement versus velocity memory}\label{sec5}

%------------------------------------------------------------------------------------------  
\subsection{Generalities}\label{sec2.1}

We now have all the elements to discuss the memory effects depending on the form of the wave-profile. To simplify the discussion, we focus on wave-profiles which satisfy at early time $u_0$
\be
\label{early}
A_{ij}(u_0) = \delta_{ij}\,.
\ee
Notice that this \textit{does not} that imply $\dot{A}_{ij}(u_0) =0$ at $u_{0}$.  Consider a test particle which is either massive, i.e. $\epsilon = -1$, or massless, i.e. $\epsilon =0$, and let us analyze its motion in the 2d plane orthogonal to the direction of the propagation of the wave for a general $\epsilon$. The motion in this 2d space does not depend on the value of $\epsilon$.
Plugging the solution of the geodesic motion (\ref{xtraj}) into Eq.~(\ref{FNC3}), one finds that the position and the velocity of the test particle are given in Brinkmann coordinates by
\begin{align}
\label{XFNC}
X^A & =   E^A{}_i \; \left( H^{ij} p_j - \B^i\right)\,, \\
\label{XFNCdot}
\dot{X}^A & =  \dot{E}^A{}_i  \left( H^{ij}p_j   - \B^i\right) + E^A{}_i A^{ij} p_j  \,.
\end{align}
There are parametrized by the two 2d vectors $(p_i, \B^i)$, which are constants of motion related to the initial conditions we have introduced in the previous section.  From Eqs.~(\ref{XFNC}) and~(\ref{XFNCdot}), the initial position and velocity of the test particle in the 2d transverse plane are related to the constants $(\B^i,p_j)$ as
\be
X^A (u_0)= -\delta^A{}_i \B^i\,, \qquad \dot{X}^A (u_0) = \delta^A{}_i\left( \delta^{ij} p_j -  \dot{\A}_{i}(u_0) \B^i\right)\,.
\ee
Note that $\dot X^A$ represents the 2d projection of the relative tangent vector, which is different from the 2d-projection of the 3-velocity, $\dot X^A/\dot T$. 

Now, consider a couple of test  particles and choose one of them as the reference geodesic, i.e. set $X^{\mu}:= \bar{X}^{\mu}$ and let us denote the geodesic of the other test particle by $Y^{\mu}(\tau)$. Both satisfy Eqs.~(\ref{xtraj}) and~(\ref{vtraj}) with initial conditions~(\ref{chargeref}) for the reference geodesic and Eqs.~(\ref{e.bxO},\ref{e.bvO}) for the second one. Its charges $p^{(Y)}_i$ and $\B^i_{(Y)}$ fix its momentum and its position in the 2d transverse plane. It follows that their relative displacement in the 2d transverse plane, that is the 2d distance between the two geodesics in Fermi coordinates, is defined as
 \be
 \zeta^A \equiv Y^A - \bar{X}^A\,,
 \ee
 so that $\dot{\zeta}^A$ represents their 2d relative velocity in the transverse plane.  Since $\bar{p}_i=0$ and $\bar{\B}_i =0$ for the reference geodesic, one can directly reads the physical distance in the Fermi coordinates from the formula (\ref{XFNC} -\ref{XFNCdot}). Solving the parallel transport equation for $E^A{}_i$ and imposing the restriction (\ref{early}), one finds that
 \begin{align}
 E^A{}_i  = \A_{i}(u) E^A{}_i (u_0) \,,\qquad  \dot{E}^A{}_i  = \dot{\A}_{i}(u) E^A{}_i (u_0)\,.
 \end{align}
Using these expressions, the physical distance in the Fermi coordinates between the test particle following the trajectory $Y^{\mu}$ and the reference geodesic $\bar{X}^{\mu}$ is given by
\be
\zeta^A = \zeta^i(u) E^A{}_i (u_0) 
\ee
where
\be\label{e.f4}
\zeta^i(u) = {\cal A}_i(u)\left[ H^{ii}(u_0,u) p_i - \B^i \right] \,.
\ee
Notice that because the matrix are diagonal, there are no summation on the indice $i$.  The Souriau matrix can be written as
\be
\label{Hint}
H^{ij} \equiv H^{ii} \delta^{ij} \quad\hbox{with}\quad H^{ii}( u_0,u)=\int_{u_0}^u \frac{\dd v}{{\cal A}^2_i(v)}.
\ee
Using $A^{ii} = A^{-1}_{ii}$ and the expression for $H^{ii}$, one can derive simple expressions for the relative velocity and acceleration between the two trajectories, which read
\begin{align}
\label{e.f5}
\dot \zeta^i(u) & = \frac{\dot {\cal A}_i(u)}{{\cal A}_i(u)} \zeta^i(u)  +\frac{p_i}{{\cal A}_i(u)}\,, \\
\label{e.f6}
\ddot \zeta^i(u) & =  \frac{\ddot {\cal A}_i(u)}{{\cal A}_i(u)} \zeta^i(u).
\end{align}
Note that Eq.~(\ref{e.f5}) rewrites as
\be\label{rrel}
 p_i = \dot\zeta_i\A_i -\dot\A_i\zeta_i
\ee
without summation on $i$, at all $u$. This relation will turn to be useful in the following.\\

\subsection{Classification of the memory effects}\label{sec2.2}

We can now use these expressions to classify the different conditions for having a displacement or a velocity memory effect. We are interested in the situations for which asymptotycally, i.e. for $u<u_0$ and $u>u_f$, the relative distance is non-zero and the relative acceleration vanishes, which read
\be
\label{COND1}
\zeta^i \neq 0 \qquad \text{and} \qquad \ddot{\zeta}^i = 0 \,.
\ee

We shall distinguish three different cases of memory effects that satisfy the above condition.
\begin{itemize}
\item \textbf{Velocity Memory (VM)}: When the relative velocity in the two asymptotic regions $u<u_0$ and $u>u_f$ satisfies
\be
\Delta \dot{\zeta} = \dot{\zeta}_f - \dot{\zeta}_0 \neq 0
\ee
the dynamics exhibits a {\em constant velocity memory effect}, i.e. the passage of the gravitational wave induces a constant shift on the asymptotic value of the relative velocity.
\item \textbf{Vanishing Velocity Memory (VM0)}: When the shift between the asymptotic relative velocities satisfies
\be\label{e.condDMpulse}
\Delta \dot{\zeta} = \dot{\zeta}_f - \dot{\zeta}_0 =0
\ee
there is clearly no velocity memory effect since the relative velocity is the same in the two asymptotic regions. Yet, as we shall see, the wave can still have some interesting effects on the couple of test particles.
\item \textbf{Displacement Memory (DM)}: this situation is characterized by the fact that in the asymptotic future, the relative velocity vanishes, i.e. the solution is such that
\be\label{e.condDMgen}
\dot{\zeta}_f =0\,.
\ee
This means that the two test particles settle at a constant distance for $u>u_f$ independently of any condition on $\dot{\zeta}_0$. In particular, the case where $\dot{\zeta}_0 \neq 0$ corresponds to a configuration where the two particles have a relative velocity for $u<u_0$ and in which the gravitational wave cancels this initial relative motion so that the particles settle at a constant relative distance.  
\end{itemize}

Thanks to results of the Section~\ref{sec2.1} we have all the elements to analyze the conditions leading to three cases: VM, VM0 or DM. To that goals, one needs to distinguish two kind of profiles: the pulse and step wave-profiles, for which the asymptotic behaviors at early and late times differ.

%------------------------------------------------------------------------------------------  
\subsection{Pulse profiles}

A pulse profile is characterized by the property that the Brinkmann profile $H_{+}(u)$ vanishes in both asymptotic regimes, i.e. for $u < u_0$ and $u > u_f$. It follows that the asymptotic behaviors of $\lbrace \A_1(u), \A_2(u)\rbrace$,  obtained by solving Eq.~(\ref{wavBrink1}), automatically satisfy
\be\label{COND0}
\ddot{\A}_i = 0\,.
\ee
Both modes behave identically and independently so that we temporarily drop the indices $i$ in $\A_i$ and $\zeta^i$. We now analyze their asymptotic behavior.

First, in the remote past, where $u < u_0$, it is obvious from Eqs.~(\ref{COND1}) and (\ref{COND0}) that $\A(u)$ and $\zeta(u)$ behave as
\begin{align}
\A (u) & = \dot{\A}_0 (u-u_0) + \A_0 \,, \label{e.Au0} \\
\zeta(u) & = \dot{\zeta}_0 (u-u_0) + \zeta_0\, \label{e.zetau0}
\end{align}
with $( \A_0,\dot{\A}_0 ,\zeta_0,\dot{\zeta}_0)$ constant. This solution is indeed consistent with Eq.~(\ref{wavBrink1}) for the asymptotic form of a  pulse profile. Hence, in this regime, the integral (\ref{Hint}) becomes
\be
H(u,u_0) = \int^u_{u_0} \frac{1}{\A^2(u)} = \frac{u-u_0}{\A_0 \A(u)} \qquad \text{such that} \qquad \A H(u,u_0) = \frac{u-u_0}{\A_0}\,,
\ee
which remains well-defined even if ${\cal A}$ crosses zero (see Appendix~\ref{appC} for a general discussion on this issue). From Eq.~(\ref{rrel}), we have that
\be\label{e.CI0a}
\dot{\zeta}_0 {\A_0} - \zeta_0 \dot{\A}_0 = p\,.
\ee
To be consistent, the relation (\ref{e.f4}) shall hold for all $u<u_0$, which implies that the constants of motion $(p,\B)$ are related to the initial conditions in $u_0$ by
\begin{align}\label{e.CI0b}
\zeta_0 = -  \B\A_0 \,, \qquad \dot{\zeta}_0 = \frac{p}{\A_0} -  \B \dot{\A}_0\,
\end{align}
which is indeed consistent with Eq.~(\ref{e.CI0a}). This fully determined the evolution of the spacetime geometry and of the geodesic before the pulse.

\begin{figure}[htb]
 	\centering
 	\includegraphics[width=0.3\textwidth]{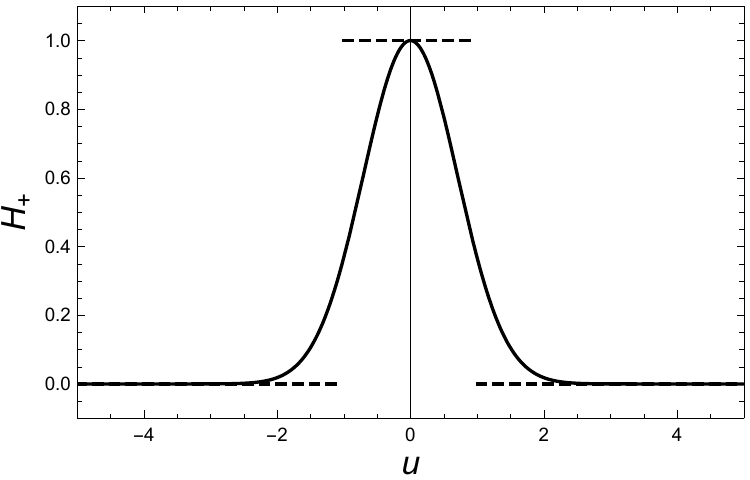} \includegraphics[width=0.3\textwidth]{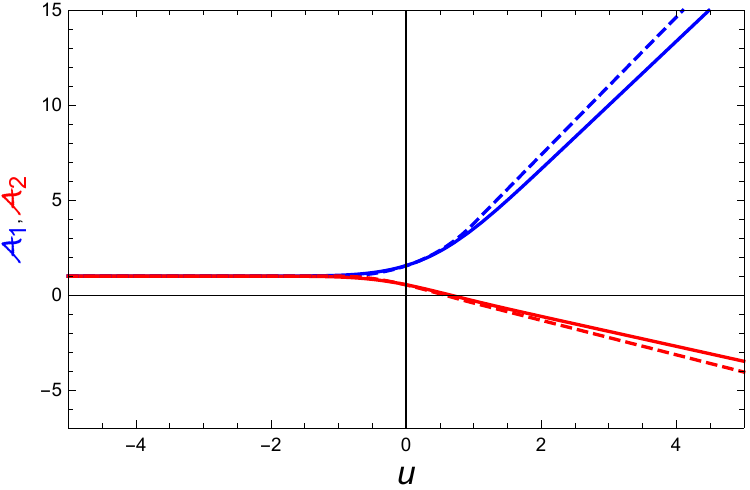} \includegraphics[width=0.3\textwidth]{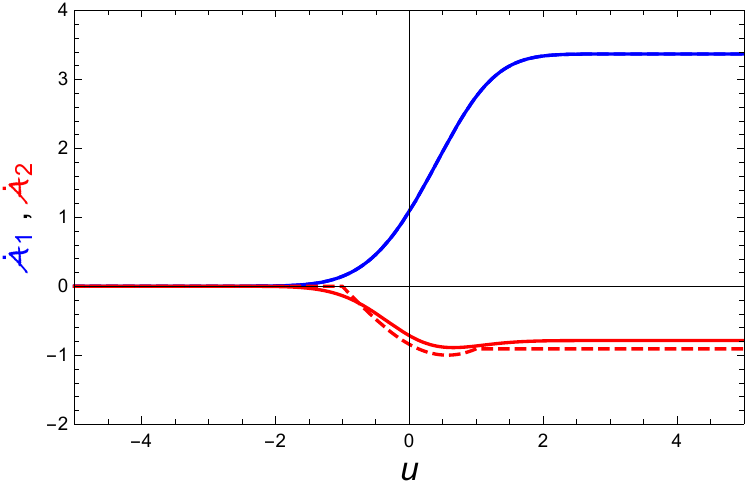}
	\vspace{0.cm}
 	\caption{{\it Left}: Profile of the pulse either as $H_+=\hbox{e}^{- u^2}$ (Solid) or a top-hat (Dashed). {\it Middle}:  ${\cal A}_1$ (blue) and ${\cal A}_2$ (red) solution of Eq.~(\ref{wavBrink1}) with initial conditions $A_{ij}(u_0)=\delta_{ij}$. {\it Right}: Associated $\dot{\cal A}_1$ (blue) and $\dot{\cal A}_2$ (red). }
 	\label{fig:1}
 	\vspace{0.25cm}
 \end{figure} 

\begin{figure}[htb]
 	\centering
 	\includegraphics[width=0.4\textwidth]{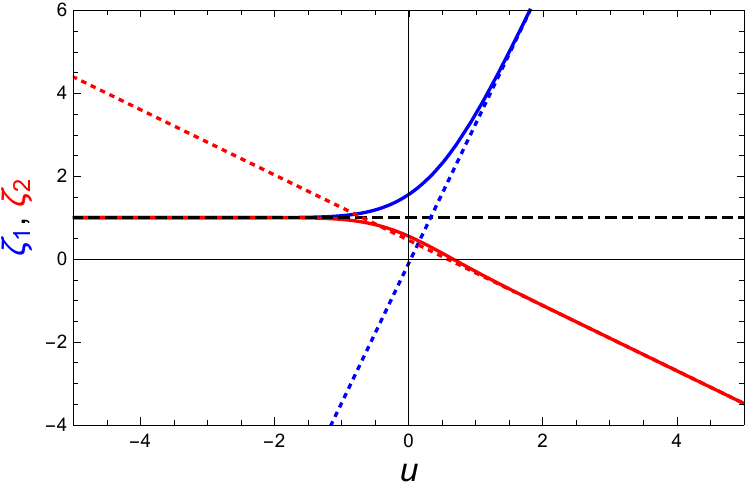} \includegraphics[width=0.4\textwidth]{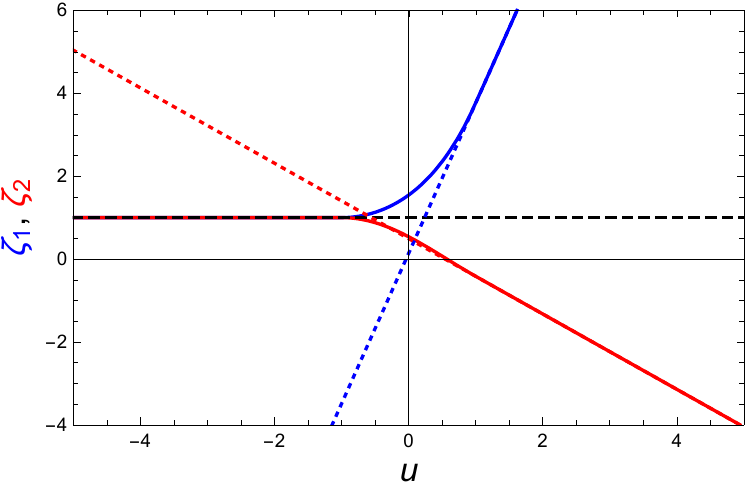}
	
	\includegraphics[width=0.4\textwidth]{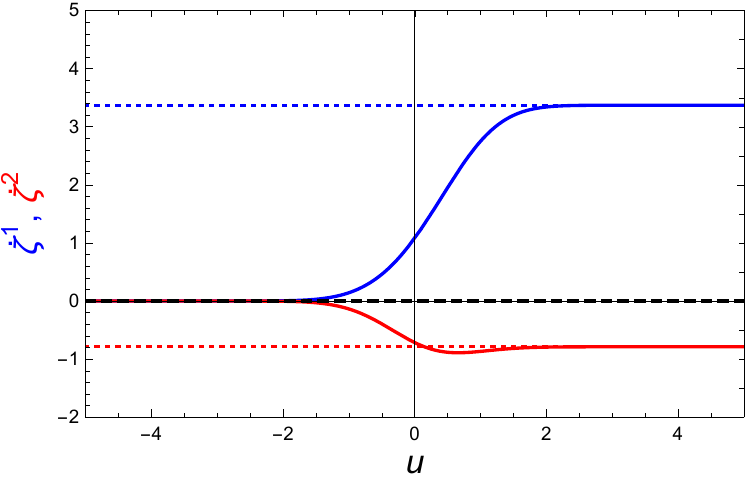} \includegraphics[width=0.4\textwidth]{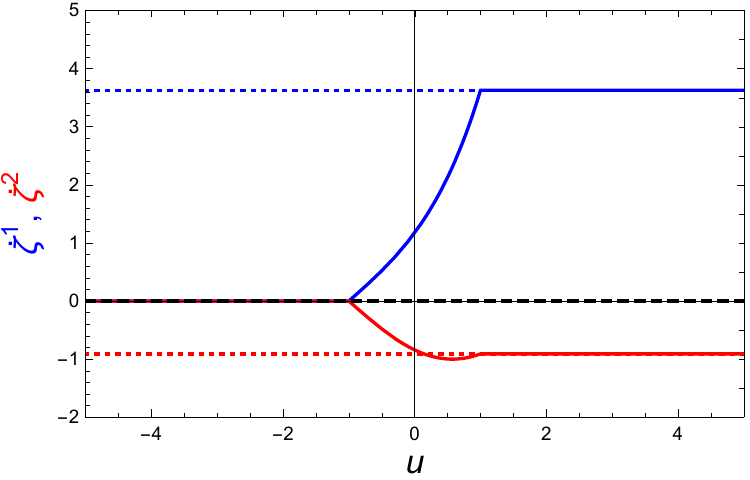}
	\vspace{-0.75cm}\smallskip \smallskip
 	\caption{The profiles of $\zeta_1$ and $\zeta_2$ (top) and $\dot\zeta_1$ and $\dot\zeta_2$ (bottom) for  $H_+=\hbox{e}^{- u^2}$ (left) and a top-hat (right) assume $p_i=(0,0)$ and $\B_i=(-1,-1)$. The dashed line corresponds to the analytic asymptotic linear motion for $u<u_0$ given by Eq.~(\ref{e.zetau0}) and the dotted lines for $u>u_f$ given by Eq.~(\ref{e.zetauf}). This illiustrates the generic existence of a bounded VM for pulse profiles.}
 	\label{fig:2}
 	\vspace{-0cm}
 \end{figure}

Then, at late time when $u > u_f$, $\A(u)$ and $\zeta(u)$ also behave, from Eqs.~(\ref{COND1}) and (\ref{COND0}), as
\begin{align}
\A (u) & = \dot{\A}_f (u-u_f) + \A_f\,,  \label{e.Auf} \\
\zeta(u) & = \dot{\zeta}_f (u-u_f) + \zeta_f\,. \label{e.zetauf}
\end{align}
In this regime, the integral (\ref{Hint}) takes the form
\be
H(u_0,u) = \int^{u_f}_{u_0} \frac{1}{\A^2(u)} + \int^{u}_{u_f} \frac{1}{\A^2(u)} = H_{0f}  + \frac{u-u_f}{\A_f \A(u)} \,.
\ee
The constant $H_{0f}\equiv H(u_0,u_f)$ encodes the behavior of the wave-profile $H_+$ between the two asymptotic regimes.  Following the same steps as for $u<u_0$, Eq.~(\ref{rrel}) implies  that
\be\label{e.CI0c}
\dot{\zeta}_f {\A_f} - \zeta_f \dot{\A}_f = p\,,
\ee
and the relation (\ref{e.f4}) to implies that  $(\A_f, \zeta_f, \dot{\A}_f, \dot{\zeta}_f)$ are related to the constants of motion $(\B,p)$ by
\begin{align}
\zeta_f & = \A_f (H_{0f} p - \B) \\
\label{relspeedf}
\dot{\zeta}_f & =   \frac{p}{\A_f} + \dot{\A}_f (H_{0f} p - \B)\,,
\end{align}
which are easily checked to be consistent with Eq.~(\ref{e.CI0b}). Such a pulse solution is illustrated for two pulse profiles on Fig.~\ref{fig:1} for the spacetime geometry and on Fig.~\ref{fig:2} for the evolution of $(\zeta_1,\zeta_2)$ and their derivative. They compare the full numerical integration to the asymptotic behaviors~(\ref{e.zetau0}) and~(\ref{e.zetauf}) that are obtained analytically. Note also that $\zeta_2$ is well-behaved even if ${\cal A}_2$ crosses zero, as discussed in Appendix~\ref{appC}.\\

Since Eq.~(\ref{e.CI0b},\ref{relspeedf}) rewrites as simple system for the constants of motion $(p,\B)$,
\begin{align}
\left\lbrace
\begin{array}{l}
 - \dot{\A}_0   \B +   \frac{1}{\A_0} p= \dot{\zeta}_0  \\
- \dot \A_f \B + \left( \frac{1}{\A_f} +\dot{\A}_f H_{0f} \right) p = \dot{\zeta}_f
\end{array}
\right. \,,
\end{align}
one can always find a unique solution $(p,\B)$ so that given a pulse profile, there exists a solution that interpolate between $\dot\zeta_0$ and $\dot\zeta_f$ provided $\dot{\A}_0/\A_f- \dot \A_f/\A_0+\dot{\A}_0\dot{\A}_f H_{0f} \not=0$.  It follows that these solutions in the two asymptotic regimes allow one to easily analyze the conditions for the different memory effects to occur. We restrict to the 3 cases defined in \S~\ref{sec2.2}.

A \textbf{VM} occurs under the condition $ \dot{\zeta}_f \neq  \dot{\zeta}_0$. Since pulses enjoy a constant asymptotic velocities both for $u<u_0$ and $u>u_f$, they generically lead to a constant VM whatever the profile $H_+(u)$ and the constants of motions $(\B,p)$ or, similarly, whatever the initial conditions $(\zeta_0,\dot\zeta_0)$. This is a generic properties of pulse illustrated on Fig.~\ref{fig:2}.

\begin{figure}[h]
 	\centering
 	\includegraphics[width=0.3\textwidth]{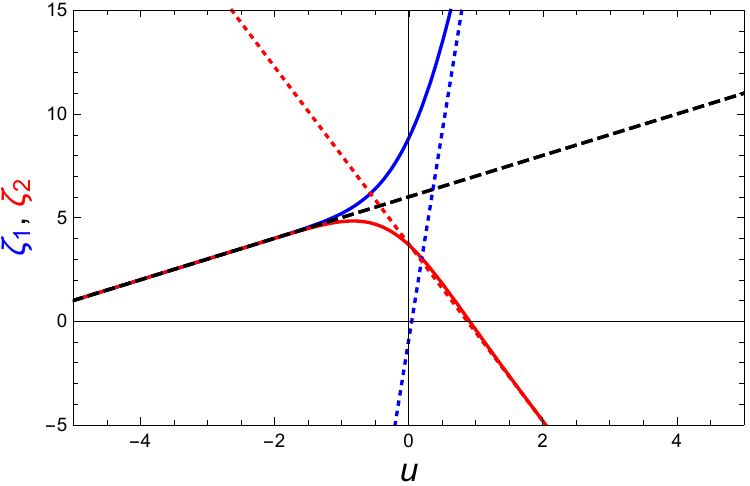}\includegraphics[width=0.3\textwidth]{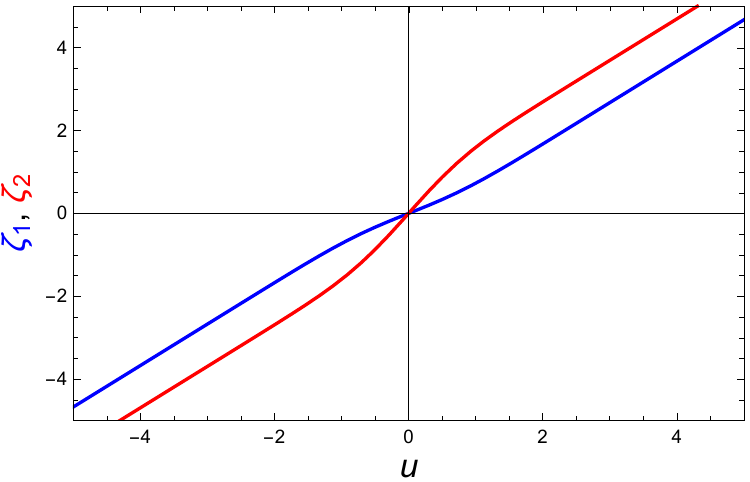} \includegraphics[width=0.3\textwidth]{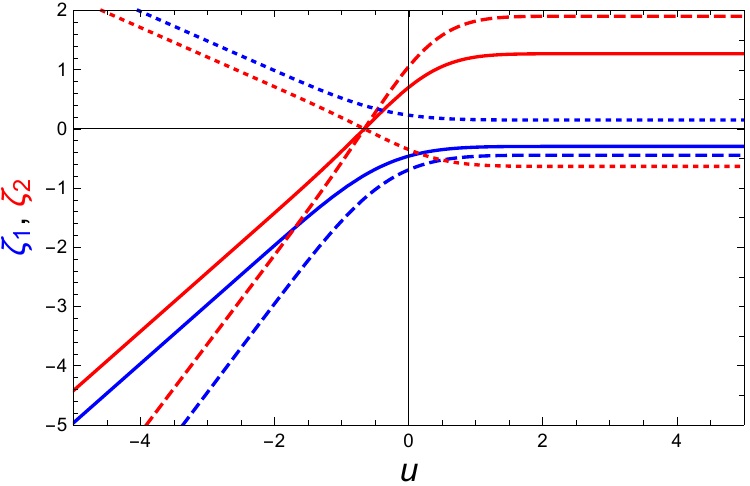} 
	
	 \includegraphics[width=0.31\textwidth]{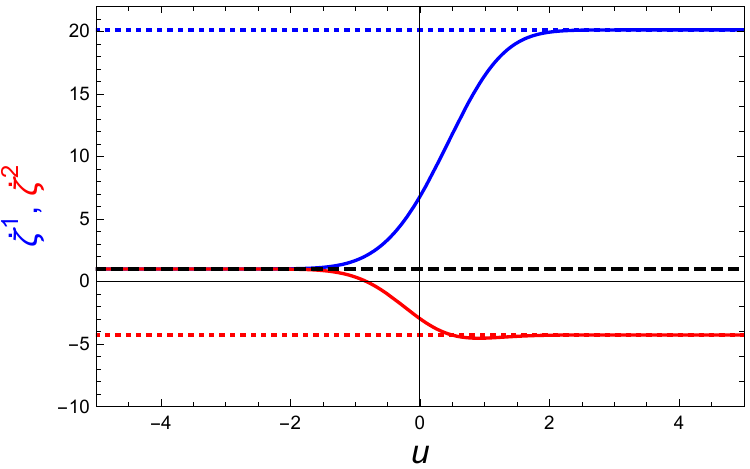}\includegraphics[width=0.3\textwidth]{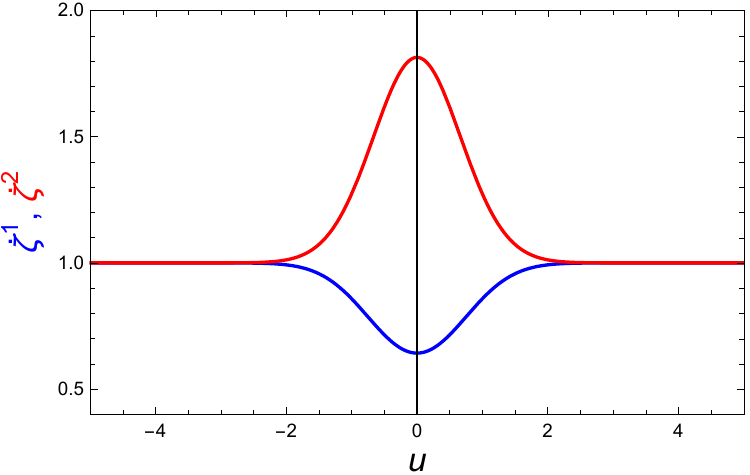}\includegraphics[width=0.3\textwidth]{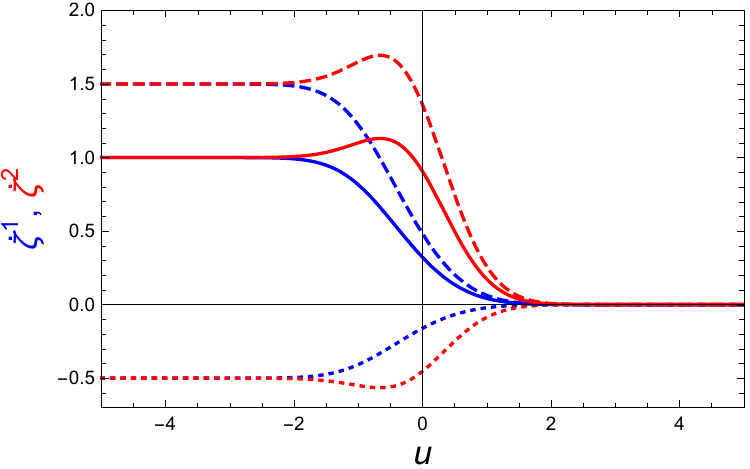}
	\vspace{-0cm}
 	\caption{Profiles of the  relative displacement $(\zeta_1,\zeta_2)$ (upper line) and relative velocity $(\dot\zeta_1,\dot\zeta_2)$ (lower line) for  $H_+=\hbox{e}^{- u^2}$ with initial conditions that ensures $\dot{\cal A}_f-\dot{\cal A}_0\not=0$. On the {\it ledt} panels one assumes $p_i=(1,1)$ and $\B_i=(-1,-1)$ showing a clear non vanishing VM, so that $\dot\zeta_f\not=\dot\zeta_0$. On the {\it middle} column, $B_i$ is tuned as in Eq.~(\ref{e.Cpulse1}) to get a VM0, i.e. a vanishing VM for which $\dot\zeta_f=\dot\zeta_0$, which indeed requires $p_i\not=(0,0)$. On the {\it right} column, $B_i$ is tuned as in Eq.~(\ref{condDM}) so that the final relative velocity vanishes while $\dot\zeta_0=p/\A_0-\B\dot\A_0$ so that we get a pure constant DM, i.e. a pure relative separation between the two particles after the passage of the wave, for the 3 cases $p_i=1$ (Solid), $p_i=1.5$ (Dahed) and $p_i=-0.5$ (Dotted).}
 	\label{fig:3}
 	\vspace{-0.3cm}
 \end{figure}

A \textbf{VM0} occurs in the special cases in which the change of the relative velocity vanish, i.e. when the condition~(\ref{e.condDMpulse}) holds. This condition, i.e. $\dot\zeta_f=\dot\zeta_0$, translates into
\begin{align}\label{e.Cpulse0}
(\dot{\A}_f - \dot{\A}_0)\B = - p \left[ \frac{\A_f - \A_0}{\A_f \A_0} - \dot{\A}_f H_{0f}\right]\,.
\end{align}
This equality can be achieved in different ways.
\begin{enumerate}
\item If $\dot \A_f-\dot \A_0\not=0$, then a VM0 occurs if only if the momenta $p$ and the initial position $\B$ are related by
\begin{align}\label{e.Cpulse1}
\B = - p \, \frac{\frac{\A_f - \A_0}{\A_f \A_0} - \dot{\A}_f H_{0f}}{\dot{\A}_f - \dot{\A}_0}\,.
\end{align}
Since $\B\not=0$, the VM0 is possible if and only if $p\not=0$ and only for some specific geodesics satisfying (\ref{e.Cpulse1}). This case is illustrated in Fig~\ref{fig:3}. It shows that even if there is no change in the asymptotic relative velocity, the wave induces a switch of the positions of the two particles.
\item If $\dot \A_f-\dot \A_0=0$ then,
\begin{itemize}
\item whatever the wave profile and whatever $\B$, one has a VM0 if
\be
p=0\,.
\ee
\item if $p\not=0$, then the wave profile shall satisfy 
\be\label{e.Cpulse2}
H_{0f} = \frac{\A_f - \A_0}{\A_0 \A_f \dot{\A}_f }\,,
\ee
which is a non-trivial condition between ${\cal A}$ in the two asymptotic regions and the shape $H_+(u)$ of the pulse through $H_{0f}$.
\end{itemize}
\end{enumerate}

Finally, a \textbf{DM} occurs when the condition~(\ref{e.condDMgen}) holds, i.e. when $\dot{\zeta}_f =0$. Using Eq.~(\ref{relspeedf}), this translates into a specific tuning between the two constants of motion,
\be
\label{condDM}
\dot{\A}_f \B =  p \left( \frac{1}{\A_f} + \dot{\A}_f H_{0f}\right)\,.
\ee
This condition can be satisfied in several ways. 
\begin{enumerate}
\item If $\dot \A_f \not=0$, then a pure DM requires $p$ and $\B$ to be related by
\begin{align}
\label{condDMm}
\B = p \, \frac{1 + \A_f \dot{\A}_f H_{0f}}{\A_f \dot{\A}_f}\,.
\end{align}
Since $\B\not=0$, the DM is possible if and only if $p\not=0$.
\item If $\dot \A_f=0$ then,
\begin{itemize}
\item whatever the wave profile and whatever $\B$, one has a DM if
\be
p=0\,.
\ee
\item if $p\not=0$, then the wave profile shall satisfy 
\be\label{e.Cpulse2}
H_{0f} = - \frac{1}{ \A_f \dot{\A}_f }\,,
\ee
which is a non-trivial condition between ${\cal A}$ in the asymptotic after zone and the shape $H_+(u)$ of the pulse through $H_{0f}$.
\end{itemize}
\end{enumerate}
This is illustrated on the bottom line of Fig.~\ref{fig:3} which shows for a given pulse profile, a generic solution (left column) with a VM, a VM0 when Eq.~(\ref{e.Cpulse1}) holds (middle column) and a pure DM if Eq.~(\ref{condDM}) is satisfied (left column).

This completes the classification of the conditions for the realization of the different memories in the case of a pulse profile. We now turn to the step profiles.
 
%------------------------------------------------------------------------------------------  
\subsection{Step profiles}

\begin{figure}[htb]
 	\centering
 	\includegraphics[width=0.3 \textwidth]{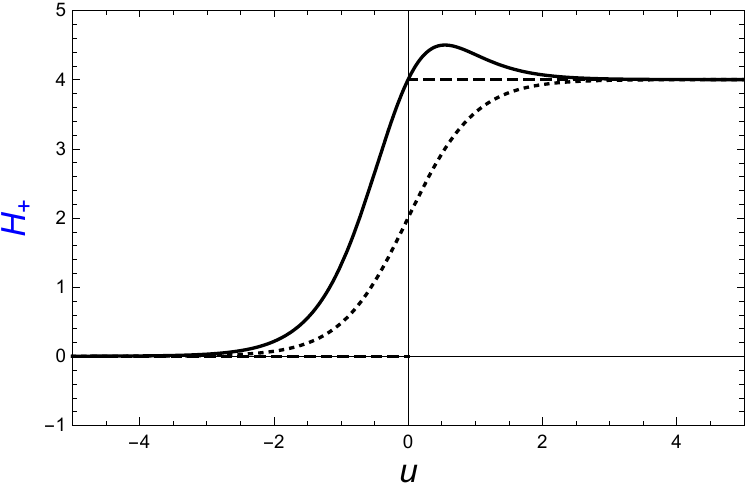} \includegraphics[width=0.3\textwidth]{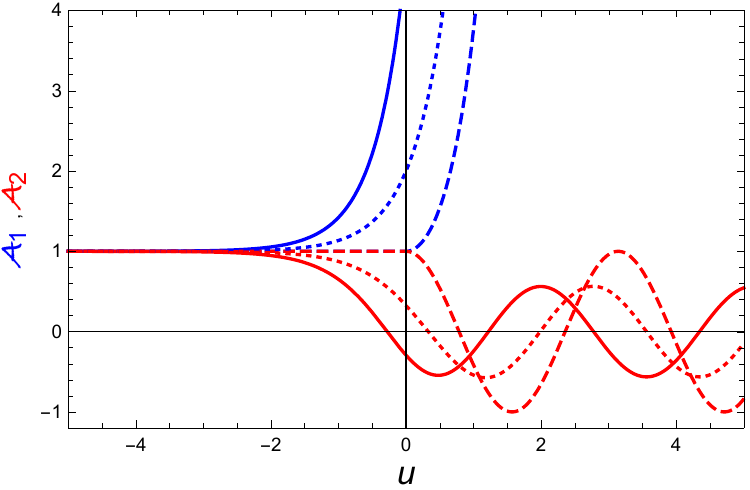} \includegraphics[width=0.3\textwidth]{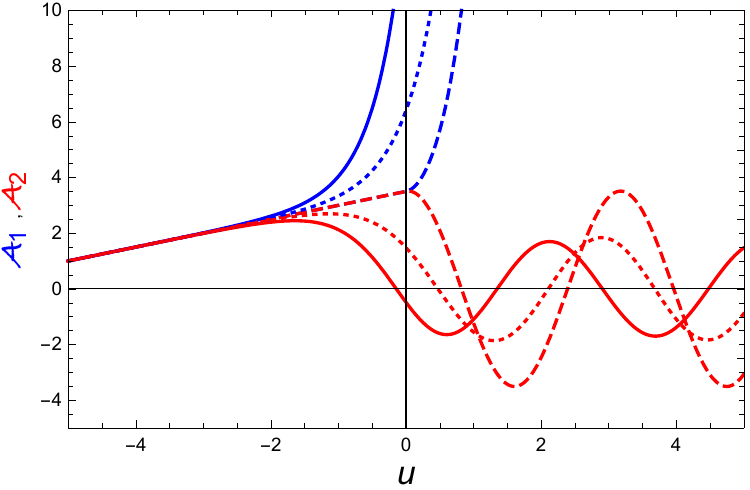}
	\vspace{-0.75cm}\smallskip \smallskip
 	\caption{{\it Left}: Profile of a step either as $H_+=4+2(1-\tanh u)\tanh u$ (Solid), $H_+=2(1+\tanh u)$ (dotted) or a Heavyside function (Dashed). {\it Middle}:  ${\cal A}_1$ (Blue) and ${\cal A}_2$ (Red) solution of Eq.~(\ref{wavBrink1}) with initial conditions $A_{ij}(u_0)=\delta_{ij}$ for the three profiles. {\it Right}: One assumes $\dot A_{ij}(u_0) =\frac{1}{2}\delta_{ij}$. This illustrates that while their functional forms in the asymptotic regions are identical the specific form of the profile affects the relation between $({\cal A}_f, \dot{\cal A}_f)$ and $({\cal A}_0, \dot{\cal A}_0)$ as well as the value of $H_{0f}$.}
 	\label{fig:4}
 	\vspace{-0cm}
 \end{figure}

Now, we consider the second family of wave profiles. A {\em step} profile is characterized by the property that the Brinkmann profile $H_{+}(u)$ is constant and non-vanishing in both asymptotic regimes, i.e. $H_+\rightarrow \lambda_{0,f}\not=0$ respectively for $u < u_0$ and $u > u_f$. Let us stress that while pulse profiles have been studied at length in the literature, step profiles have been largely ignored in the context of vacuum gravitational plane wave. They are nevertheless relevant examples as they appear to be interesting models of the non-oscillating contribution of the strain in the perturbative approach. See for example the discussion in Ref. \cite{Favata:2011qi}.  In the two asymptotic regions,  $\lbrace \A_1(u), \A_2(u)\rbrace$  are obtained by solving (\ref{wavBrink1})  and automatically satisfy
\be\label{COND0step}
\ddot{\A}_1 = \lambda_{0,f}^2 \A_1, \qquad \ddot{\A}_2 =- \lambda_{0,f}^2 \A_2
\ee
We assume $\lambda_{0,f}>0$ without loss of generality. Examples of step profiles and of geometries $(\A_1,\A_2)$ are depicted on Fig.~\ref{fig:4}.

First, in  the past region $u < u_0$ in which that $H_+ = \lambda^2_{0} >0$, the solutions of the equations of evolution~(\ref{wavBrink1}) are  easily obtained as
\begin{align}
\A_1 (u) & =  \A_0 \cosh\left[ \lambda_0 (u - u_0) \right] + \frac{\dot{\A}_0}{\lambda_0}\sinh   \left[  \lambda_0 (u - u_0) \right] \\
\A_2 (u) & =  \A_0 \cos \left[ \lambda_0 (u - u_0)  \right]  + \frac{\dot{\A}_0}{\lambda_0}\sin  \left[  \lambda_0 (u - u_0)\right] \,.
\end{align}
Contrary to the pulse profile, we shall discuss the asymptotic memory effects, i.e. between the regions $(-)$ and $(+)$ for $\rightarrow\pm\infty$ respectively. It is clear that
\be\label{e.reg-}
\A^{-}_1 \sim \frac{1}{2} \left( \A_0 - \frac{\dot\A_0}{\lambda_0}\right) \hbox{e}^{-\lambda_0 (u-u_0)}\, \qquad
\dot{\A}^{-}_1 =-\lambda_0A^{-}_1\, \qquad
\ddot{\A}^{-}_1 =\lambda_0^2A^{-}_1\,.
\ee
Hence, obviously,  unless $\A_0- \frac{\dot{\A}_0}{\lambda_0} =0$, $\A_1(u)$ is exponentially diverging  when  $u \ll u_0$ and $\A_2(u)$ is either oscillating or strictly vanishing. In this regime, the Souriau matrix (\ref{Hint}) becomes
\begin{align}
H^{ii} (u_0,u) = \int^u_{u_0} \frac{\rd v}{\A_i^2(v)} =
\frac{1}{\lambda_0 \A_i(u) \A_i(u_0)} \times \left\lbrace
\begin{array}{ll}
\sinh[\lambda_0(u-u_0)] & i=1  \\
\;\;\sin[\lambda_0(u-u_0)] & i=2 
\end{array} \right. \,.
\end{align}
The solutions for the mode $\A_2$ (and hence $\zeta_2$) are obtained  from the expression of the mode $\A_1$ (and $\zeta_1$) by the transformation $\lambda_0 \rightarrow i \lambda_0$. Again, since both modes behave identically and independently, this allows us to drop the indices $i$ and work with $\A=\A_1$. The expressions for the mode 2 can then be easily obtained but one has to keep in mind the difference of behavior since $\A_1$ behaves exponentially while $\A_2$ remains bounded as it oscillates and crosses 0 periodically. Again we stress that $H(u_0,u)$ remains well-defined even if ${\cal A}$ crosses 0 (see Appendix~\ref{appC} for a general discussion on this issue). 

Since Eq.~(\ref{e.f5}) rewrites $\dot\zeta{\cal A}-\dot{\cal A}\zeta=p$ for all $u$, it again implies that
\be\label{e.CI0stepa}
\dot{\zeta}_0 {\A_0} - \zeta_0 \dot{\A}_0 = p\,.
\ee
Then, the relation (\ref{e.f4}) gives
\begin{align}\label{e.zetastep0}
\zeta(u) & = p  \frac{ \sinh{[\lambda_0(u-u_0)]}}{\lambda_0 \A_0} -  \B\A(u)\,.
\end{align}
Since it  shall hold for all $u<u_0$, it implies that the constants of motion $(p,\B)$ are related to the initial conditions in $u_0$ by
\begin{align}\label{e.condICstep0}
\zeta_0 = - \B\A_0\, ,  \qquad \dot{\zeta}_0 = \frac{p}{\A_0} -  \B \dot{\A}_0\,, \qquad \ddot{\zeta}_0 = - \B  \ddot{\A}_0\, ,
\end{align}
which are indeed compatible with Eq.~(\ref{e.CI0stepa}). Interestingly they take the same form as for a pulse. 

Now, we perform the same analysis in the future asymptotic regime for $u > u_f$,  in which $H_{+} : = \lambda^2_f>0$.  It follows that $(\A_1(u), \A_2(u))$ are given by
\begin{align}
\A_1 (u) & =  \A_f \cosh\left[ \lambda_f (u - u_f) \right] + \frac{\dot{\A}_f}{\lambda_f}\sinh   \left[  \lambda_f (u - u_f) \right] \\
\A_2 (u) & =  \A_f \cos \left[ \lambda_f (u - u_f)  \right]  + \frac{\dot{\A}_f}{\lambda_f}\sin  \left[  \lambda_f (u - u_f)\right] \,.
\end{align}

\begin{figure}[htb]
 	\centering
 	\includegraphics[width=0.4\textwidth]{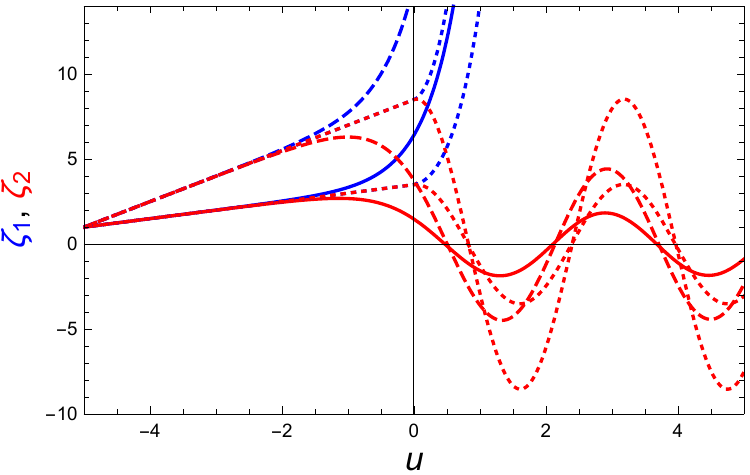} \includegraphics[width=0.4\textwidth]{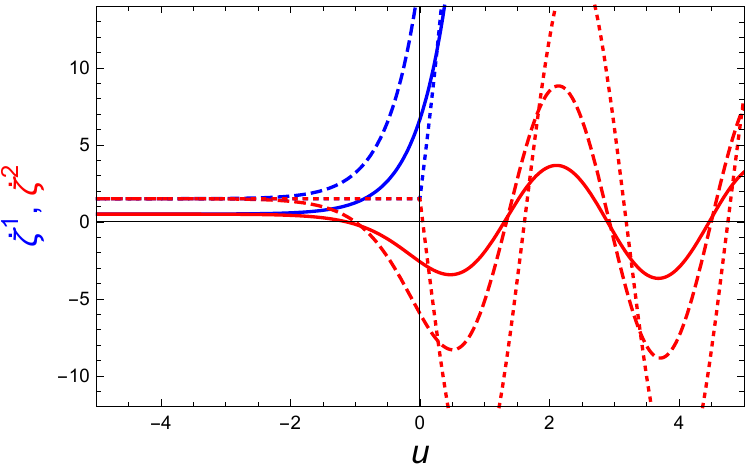}
	\vspace{-0.75cm}\smallskip \smallskip
 	\caption{The profiles of $\zeta_1$ and $\zeta_2$ (left) and $\dot\zeta_1$ and $\dot\zeta_2$ (right) for  $H_+=2(1+\tanh u)$ (Solid and Dashed)  assume $p_i=(0,0)$ (Solid) or $p_i=(1,1)$ (Dashed) and $\B_i=(-1,-1)$. The dotted line corresponds the same quantities but for $H_+$ a Heavyside function. One recovers the expected  asymptotic linear motion for $u<u_0$ and the exponential growth $u>u_f$. This illustrates the existence of a VM that is in general not bounded. Note that even if $H_+$ is discontinuous, $\zeta$ and $\dot\zeta$ remain continuous.}
 	\label{fig:5}
 	\vspace{-0cm}
 \end{figure} 
 As previously, it is clear that
\be\label{e.reg+}
\A^{+}_1 \sim \frac{1}{2} \left( \A_f + \frac{\dot\A_f}{\lambda_f}\right) \hbox{e}^{+\lambda_f (u-u_f)}\, \qquad
\dot{\A}^+_1 =\lambda_fA^{+}_1\, \qquad
\ddot{\A}^{+}_1 =\lambda_f^2A^{+}_1\,,
\ee
so that, obviously,  unless $\A_f +  \frac{\dot{\A}_f}{\lambda_f} =0$, $\A_1(u)$ is exponentially diverging when  $u \gg u_f$ and $\A_2(u)$ is either oscillating or strictly vanishing. In this regime, the Souriau matrix (\ref{Hint}) becomes
\begin{align}
H^{ii} (u_0,u) &= \int^{u_f}_{u_0} \frac{\rd v}{\A_i^2} + \int^u_{u_0} \frac{\rd v}{\A_i^2} 
  =H_{0f} +
 \frac{1}{\lambda_f \A_i(u) \A_i(u_f)}
 \times 
\left\lbrace
\begin{array}{ll}
\sinh{[\lambda_f(u-u_f)]}   & i=1\\
\;\; \sin{[\lambda_f(u-u_f)]} &i=2
\end{array} \right. \,.
\end{align}
We conclude that the evolution of $\zeta(u)$ in this regime is given by
\begin{align}\label{e.zetastepf}
\zeta(u) & = p  \left\lbrace \A(u)H_{0f} + \frac{ \sinh{[\lambda_f (u-u_f)]}}{\lambda_f \A_f}\right\rbrace -  \B\A(u)
\end{align}
These solutions, depicted on Fig.~\ref{fig:5} allow us to discuss the conditions for the different memories. The first condition to impose is that the relative acceleration vanishes, i.e. that $\ddot{\zeta} =0$ in the asymptotic regimes.  Contrary to a pulse profile, this condition is in general not met as $\ddot{\zeta}$ diverges exponentially unless it is tuned. Therefore, the first task is to compute the leading contribution to the relative acceleration and find the conditions such that it vanishes. 

In the asymptotic past, i.e. when $u < u_0$, the solution~(\ref{e.zetastep0}) implies, using Eq.~(\ref{e.condICstep0}), that
\be
\ddot\zeta(u<u_0)= \frac{p\lambda_0}{\A_0}\sinh{[\lambda_0(u-u_0)]} - \B\lambda_0^2\A(u<u_0)
\ee
so that
\begin{align}
\ddot{\zeta}_{-} =  \lim_{u\to - \infty} \ddot{\zeta} (u) & \simeq - \frac{\lambda_0}{2}   \left[ \frac{ p}{\A_0}  +  \lambda_0 \B \left( \A_0 - \frac{\dot{\A}_0}{\lambda_0}\right)  \right]  \hbox{e}^{- \lambda_0 (u-u_0)} \quad \text{when} \quad u \rightarrow - \infty\,.
\end{align}
Therefore, the only possibility to kill this divergence in relative acceleration is either to set $\lambda_0 =0$ (to be back to the pulse situation), or to enforce the condition
\be\label{Nacc0}
 (   \dot{\A}_0 - \lambda_0 \A_0 ) \B = \frac{p}{\A_0}\,.
\ee
Similarly, at late time when $u > u_f$, the leading contribution to the relative acceleration reads
\begin{align}
\ddot{\zeta}_{+} =  \lim_{u\to + \infty}  \ddot\zeta  \simeq  \frac{\lambda_f}{2}   \left[ \frac{ p}{\A_f}  +  \lambda_f ( H_{0f} p - \B) \left( \A_f + \frac{\dot{\A}_f}{\lambda_f}\right)  \right]  \hbox{e}^{\lambda_f (u-u_f)}  \quad \text{when} \quad u \rightarrow + \infty\,.
\end{align}
Demanding that $\ddot{\zeta} =0$ when $u\gg u_f$ requires that either $\lambda_f =0$ or that the condition
\be \label{Naccf}
(\dot{\A}_f + \lambda_f \A_f) \B  = p \left[ H_0f \left(\dot{\A}_f + \lambda_f \A_f \right) + \frac{1}{\A_f}\right]
\ee
holds. Indeed, the previous case of pulse profiles is recovered when $\lambda_0=\lambda_f=0$ which explains that no further conditions had to be imposed. For step profiles at least one of the two $\lambda$ shall be non-vanishing. Note also that this discussion assumes that the branching between the asymptotic regions is $\lambda_0>0\rightarrow \lambda_f>0$ in which case $\Delta\ddot\zeta_1$ diverges unless Eqs.~(\ref{Nacc0}) and~(\ref{Naccf}) hold while $\Delta\ddot\zeta_2$ is bounded and oscillating. Indeed one can also consider step profiles such that $\lambda_0<0\rightarrow \lambda_f>0$ or $\lambda_0>0\rightarrow \lambda_f<0$ in which case $\Delta\ddot\zeta_1$ oscillates in the past, resp. in the future, and for which one needs to impose~(\ref{Naccf}) in the future, resp.~(\ref{Nacc0}) in the past.\\

Our goal is not to discuss all these situations and we restrict ourselves to the more natural choice,  the one of a step profile that satisfies $\lambda_0 =0$ and $\lambda_f>0$ free, so that we only need to impose the condition~(\ref{Naccf}) in the future asymptotic region. Examples of such step profiles and the associated $(\A_1,\A_2)$ are depicted on Fig.~\ref{fig:4}. The associated profiles for the displacement and velocity for generic initial conditions are shown on Fig.~\ref{fig:5}, which illustrates the generic exponential divergence at late time for $\A_1$ and the oscillations of $\A_2$.

Once the condition (\ref{Naccf}) is imposed ensuring we have selected a configuration in which the asymptotic relative acceleration vanishes, one has to compute the asymptotic behavior of the relative velocities. Thanks to Eqs.~(\ref{e.reg-}) and~(\ref{e.reg+}) together with  Eqs.~(\ref{e.zetastep0}) and~(\ref{e.zetastepf}), one obtains
\begin{align}
\label{velfut}
\dot{\zeta}_{+}  =  \lim_{u\to + \infty} \dot{\zeta} (u) & = \frac{1}{2} \left[ \frac{p}{\A_f }+  \lambda_f ( p H_{0f} - \B) \left( \A_f + \frac{\dot{\A}_f}{\lambda_f}\right)    \right] \hbox{e}^{\lambda_f (u - u_f)}  = 0 \\
\label{velpast}
\dot{\zeta}_{-}  =  \lim_{u\to - \infty} \dot{\zeta} (u) & = \frac{1}{2} \left[   \frac{p}{\A_0 } + \B \left( \lambda_0 \A_0 -  \dot{\A}_0 \right)   \right] \hbox{e}^{\lambda_0 (u - u_0)}  = \frac{1}{2} \left[ \frac{p}{\A_0} - \B \dot{\A}_0\right] 
\end{align}
where we have used the condition (\ref{Naccf}) for the first limit and $\lambda_0=0$ for the second. We conclude that in general the asymptotic velocity does vanish in the after zone provided Eq.~(\ref{Naccf}) holds even if the asymptotic velocity is non-vanishing in the past asymptotic region. \\

From the classification of \S~\ref{sec2.2}, we thus have the following three possibilities. 

\textbf{VM}: Without any further restrictions than $\lambda_0 =0$ and the condition~(\ref{Naccf}), a wave with a step profile generically leads to a rather surprising velocity memory effect where $\dot{\zeta}_{-} \neq 0$ while $\dot{\zeta}_{+} =0$. Thus the effect of the wave is to stop the relative motion by cancelling the relative velocity between the two particles.

\textbf{VM0}: It occurs when one further has  $\dot{\zeta}_{-} =0$, which requires
\be\label{e.ttt}
\frac{p}{\A_0} = \B \dot{\A}_0 \,.
\ee
This means that a VM0 requires that $(p,\B)$ satisfy  both Eq.~(\ref{Naccf}) and ~(\ref{e.ttt}). This requires that the geometry be such that
\be\label{e.CstepVM0cond}
\frac{1}{\A_0\dot\A_0}= H_{0f} + \frac{1}{\A_f(\dot\A_f+\lambda_f\A_f)},
\ee
which is a non-trivial property of the spacetime geometry. Then, note  that one can further impose $\zeta_{+} =\zeta_{-}$ which, thanks to the condition~(\ref{rrel}), requires that 
\be
\dot{\A}_{+} = \dot{\A}_{-}
\ee
at all time. 

\textbf{DM}: A displacement memory effect requires the relative velocity to vanish in both asymptotic regimes, i.e. $\dot{\zeta}_{+} = \dot{\zeta}_{-} =0$ while the relative distance is shifted, i.e. $\zeta_{+} \neq \zeta_{-}$. Thanks to  the condition~(\ref{rrel}), this implies that one needs
\be
\label{DMM}
\dot{\A}_{+} \neq \dot{\A}_{-} \,
\ee
together with the constraint~(\ref{e.CstepVM0cond}).

\begin{figure}[htb]
 	\centering
 	\includegraphics[width=0.3\textwidth]{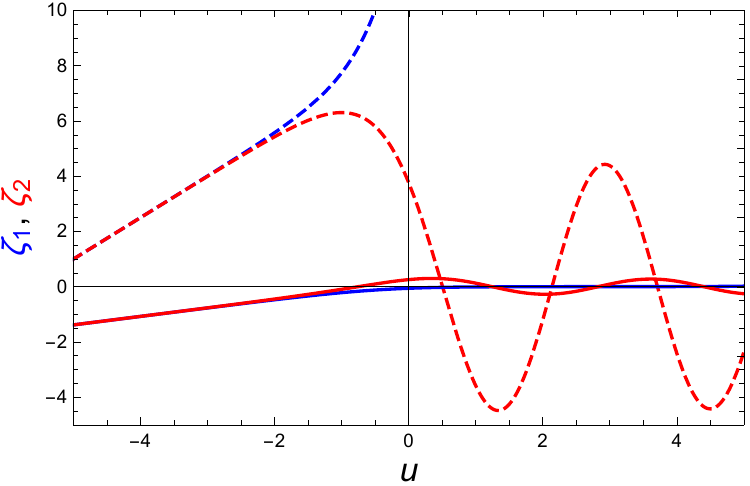} \includegraphics[width=0.3\textwidth]{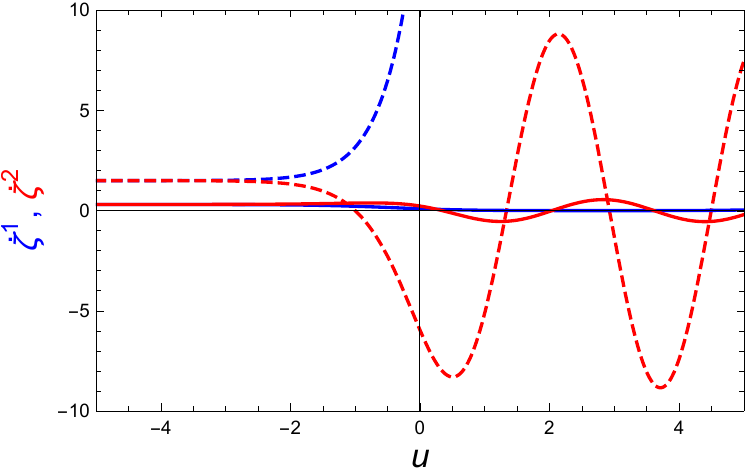}\includegraphics[width=0.3\textwidth]{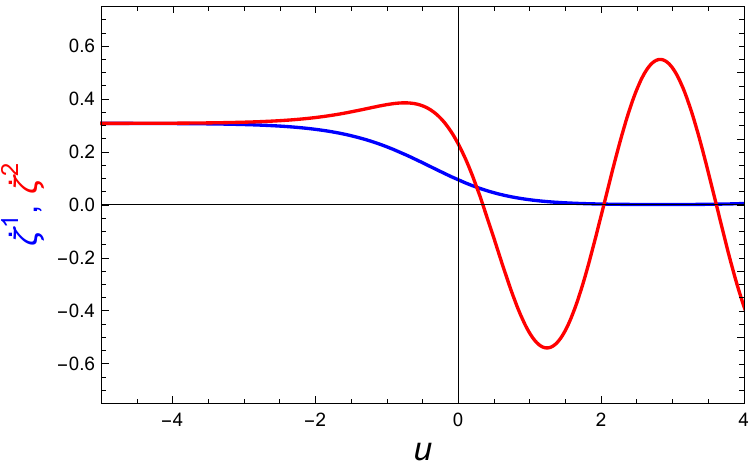}
	
	 	\includegraphics[width=0.3\textwidth]{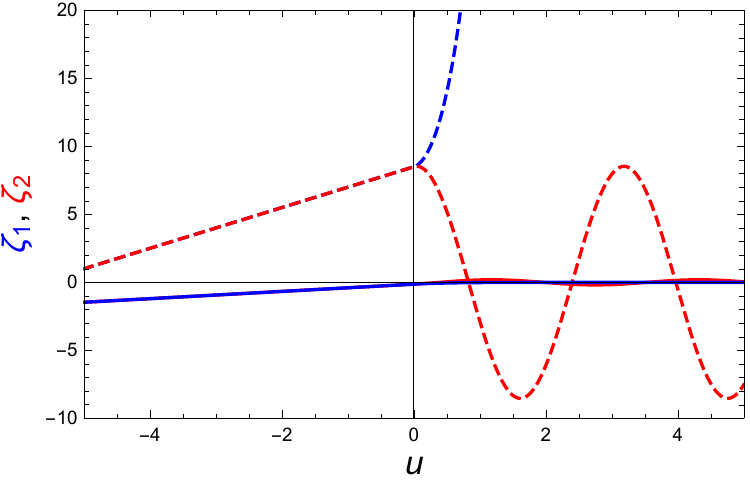} \includegraphics[width=0.3\textwidth]{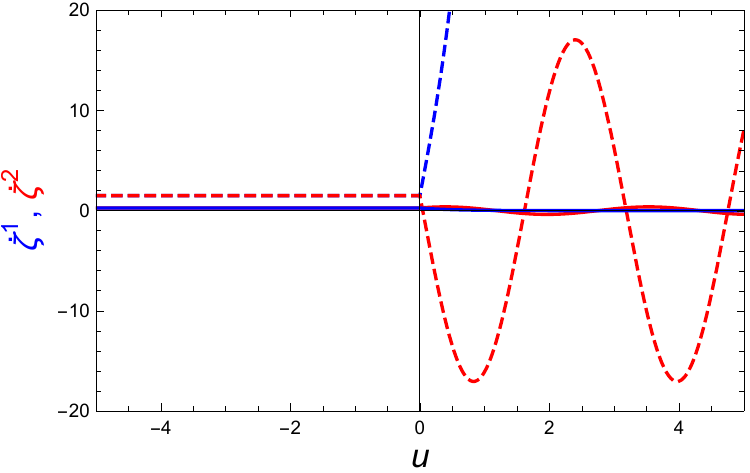}\includegraphics[width=0.3\textwidth]{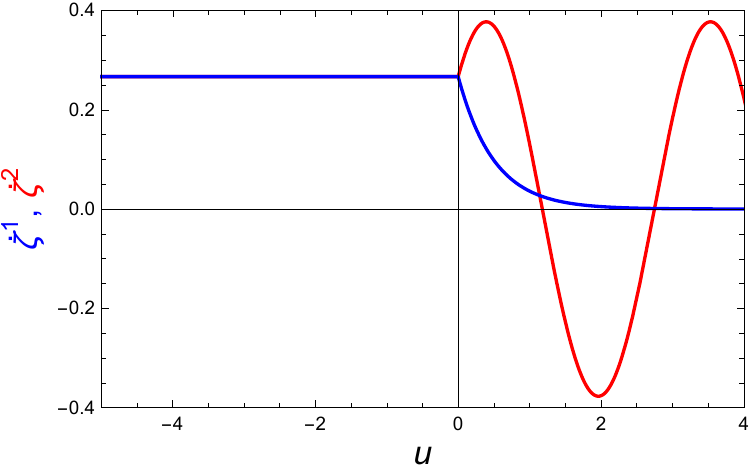}	
	\vspace{-0cm}
 	\caption{The profiles of $(\zeta_1,\zeta_2)$ ({\it left}) and $(\dot\zeta_1,\dot\zeta_2)$ ({\it middle})  for a step profile for $p_i=(1,1)$. The upper line assumes  $H_+=2(1+\tanh u)$ while the bottom line assume a Heavyside profile. On each plot, the dashed lines correspond to $\B_i=(-1,-1)$ so that the VM is not bounded at late time while the solid lines assume that $\B_i$ is determined according to Eq.~(\ref{Naccf}) so that the $\zeta_1$ enjoys a VM while $\zeta_2$ oscillates at late time. The {\it right} panel gives a rescaled view of the tuned solutions to show  the existence of a constant VM.}
 	\label{fig:6}
 	\vspace{-0.3cm}
 \end{figure}

This concludes the classification of the different memories. Figure~\ref{fig:6} compares a generic solution to the case in which the constraints~(\ref{Naccf}) is imposed, hence leading to a VM. Note that this discussions holds for the class of step profiles that interpolates from $\lambda_0=0$ to $\lambda_f>0$. To finish let us mention that if the step interpolates between $\lambda_0>0$ and $\lambda_f>0$ then the condition for a finite $\Delta\dot\zeta$ implies that $\Delta\dot\zeta=0$ but also that $\Delta\zeta=0$ asymptotically.

%====================================================================
\section{Geodesic deviation, symmetries and memories}\label{sec7}

Now, we consider the geodesic deviation equation (GDE) and show how the symmetries and their generators are related to its solutions.

To recall the construction of the GDE, consider a congruence of geodesics $X^{\mu}(\tau, \sigma)$ parametrized by $(\tau, \sigma)$. The first parameter $\tau$ stands for the proper time or affine parameter along each curve while $\sigma$ parametrizes the different curves. For curves which are very close to each other, we fix a central geodesic denoted $X^{\alpha}(\tau) := x^{\alpha}(\tau, 0)$ and follow the small deviations w.r.t. to that reference geodesic. This is done by expanding the position of the curve at fixed $\tau$ around $\sigma = 0$. The position of the nearby curves is given by
\begin{align}
\Delta X^{\mu}(\tau, \sigma) & = X^{\mu}(\tau, \sigma) -  X^{\mu}(\tau,0) \nn \\
& = \zeta^{\mu}(\tau,\sigma)  - \bar{\Gamma}^{\mu}{}_{\alpha\beta} \zeta^{\alpha} (\tau,\sigma) \zeta^{\beta}(\tau,\sigma) + \O(\sigma^3)
\end{align}
where a tilde means that the quantity is evaluated at $\sigma=0$, i.e. on the central geodesic. Here, $\zeta^{\mu}$ is the deviation vector which takes the form
\be
\label{N}
\zeta^{\mu} = \sigma N^{\mu} + % \frac{1}{2} \sigma^2 \omega^{\mu} +
 \cO(\sigma^2)\,.
\ee
Since by construction, $\bar{\Gamma}^{\mu}{}_{\alpha\beta} = 0$ in the adapted Fermi coordinates, the vector $\zeta^{\mu} \partial_{\mu}$ encodes the relative distance between the two nearby geodesics of the congruences. 
Neglecting second order corrections, one recovers the standard geodesic deviation equation
\begin{align}
\label{gde1}
 \frac{D^2 N^{\mu}}{\rd \tau^2} & = \bar{R}^{\mu}{}_{\alpha\beta \gamma} \bar{u}^{\alpha} \bar{u}^{\beta} N^{\gamma} 
\end{align}
where $D/\rd \tau = u^{\alpha} \nabla_{\alpha}$. This equation admits eight possible solutions. Let us briefly review how to decompose  and to integrate this equation when symmetries are present.

\subsection{Integrability of the geodesic deviation equation}

Decomposing the deviation in a longitudinal contribution $N_{\|}$, i.e. along $\bar{u}^{\alpha} \partial_{\alpha}$, and a contribution $N_{\perp}$ orthogonal to the $4$-velocity, such that
\be
N = N_{\|} + N_{\perp} \qquad N_{\|} = f(\tau) \bar{u}
\ee  
and  inserting this decomposition in the GDE, one finds that $f(\tau)$ is an affine function, i.e. 
\be
\label{affi}
 f(\tau)  = C_1 \tau + C_2 \,.
\ee
These two parameters are related to the freedom to rescale and translate the affine parameter (or proper time) of the reference geodesic, i.e. to the symmetry $\tau \rightarrow \tau + \epsilon$ and $\tau \rightarrow \epsilon \tau $. With these solutions, one recovers the two trivial solutions to the GDE given by $N^{\mu} = \{ \bar{u}^{\mu}, \tau \bar{u}^{\mu}\}$.  We thus focus on this six remaining solutions. The transverse contribution $N_{\perp}$ satisfies
\begin{align}
\label{gde2}
 \frac{D^2 N^{\mu}_{\perp}}{\rd \tau^2} & = \bar{R}^{\mu}{}_{\alpha\beta \gamma} \bar{u}^{\alpha} \bar{u}^{\beta} N^{\gamma}_{\perp}\,.
\end{align}
Solving this equation in a general spacetime turns out to be a difficult task. Nevertheless, exact analytical solutions can be constructed when i) the geodesic equation can be solved analytically too and ii) when the spacetime geometry admits non-trivial Killing tensors. 

Indeed, as first shown in Refs.~\cite{Caviglia0, Caviglia1, Dolan} and emphasized in Ref.~\cite{Cariglia:2018erv}, the existence of a rank-$p$ affine tensor defined as
\be
\label{SOL}
\nabla_{(\mu} K_{\alpha_1 ... \alpha_p )} = h_{(\alpha_1 ... \alpha_p )} \qquad \nabla_{\rho} h_{(\alpha_1 ... \alpha_p )} =0
\ee
implies that
\be
\label{sol}
N^{\mu} = \bar{K}^{\mu}{}_{\alpha_1 ... \alpha_p} \bar{u}^{\alpha_1} .... \bar{u}^{\alpha_p}
\ee
is a solution of Eq.~(\ref{gde1}).  In general, the space of rank-$p$ affine tensors, i.e. satisfying $\nabla_{\rho}\nabla_{(\mu} J_{\alpha_1 ... \alpha_p )} = 0$, is larger than the space of rank-$p$ Killing tensors but a full characterization does not exist. Yet, when focusing on rank-$2$ tensor, such a classification can be done. It follows that except for Petrov type N geometry, affine and Killing rank-$2$ tensors are the very same object \cite{Salmistraro:1983xr}. For Petrov type N which possesses a covariantly constant null vector $\ell_{\mu} \rd x^{\mu} = \rd \Phi$, i.e. $\nabla_{\mu} \ell_{\nu} =0$, any affine tensor can be decomposed into a Killing tensor $K_{\mu\nu}$, satisfying $\nabla_{(\alpha} K_{\mu\nu)}=0$, and a contribution coming from the covariantly constant null vector such that any rank-2 affine tensor $M_{\mu\nu}$ can be decomposed as
\be
M_{\mu\nu} = K_{\mu\nu} + \beta \Phi \ell_{(\mu} \ell_{\nu)}
\ee
thus providing a deformation of the solution based on the Killing tensor alone.

When such a geometrical object is present, it is straightforward to see that  the associated conserved charge for the geodesic motion corresponds to the projection of the deviation vector at first order, i.e. $N^{\mu}$, along the tangent vector of the geodesic congruence
\be
\label{scalprod}
M = \bar{M}^{\mu}{}_{\alpha_1 ... \alpha_p} \bar{u}^{\alpha_1} .... \bar{u}^{\alpha_p} \bar{u}_{\mu} = N^{\mu} \bar{u}_{\mu} =  f(\tau) 
\ee
When $\beta=0$, the affine tensor $M_{\mu\nu}$ reduces to the Killing tensor $K_{\mu\nu}$. From the expression (\ref{scalprod}), one could misleadingly conclude that $f(\tau)$ coincides with the Killing tensor conserved charge, such that it reduces to a constant, but it is not the case. Indeed, the evaluation of the above quantities on the reference geodesic truncates the Killing tensor and as we shall see, this scalar product is in general linear in the proper time, as expected from Eq.~(\ref{affi}). 

To summarize, if $\xi^{\mu} \partial_{\mu}$ is a Killing vector and $K_{\mu\nu} \rd x^{\mu} \rd x^{\nu}$ is a Killing tensor for the geometry, the GDE admits the following solutions
\be
N^{\mu} = \{ \bar{\xi}^{\mu} , \bar{K}^{\mu}{}_{\nu} \bar{u}^{\nu} \}
\ee
where a bar corresponds to evaluate the object on the reference geodesic, i.e. imposing (\ref{trajref}). Using this result, we now show how the different symmetries uncovered in the first section are related to the different memory effects.

\subsection{Killing vectors and transverse memory}

Consider first one of the non-trivial Killing vector (\ref{BC}) generating the Carrolian boost, say $B^{\mu} \partial_{\mu}$. In the Fermi coordinates, it is given by
\begin{align}
B^{I} \partial_I & = b_k E^A{}_i H^{ki}  \partial_A - b_k \left[ E^k{}_A   - \frac{1}{2}  \dot{A}_{ij} H^{ki} E^j{}_A \right] X^A \partial_v 
\end{align}
where $b_k$ is the symmetry parameter.
Evaluating this vector on the reference geodesic, i.e imposing (\ref{trajref}), one finds 
\begin{align}
\label{CBGDE}
\bar{B}^{I} \partial_I & = b_k E^A{}_i H^{ki} \partial_A \,.
\end{align}
Let us now check that it provides indeed a solution to the GDE. In the Fermi coordinates, the GDE reads
\begin{align}
\ddot{\zeta}_A & = R_{AUUB}   \zeta^{B} \\
& = R_{i u u j} E^{i}{}_A       E^{j}{}_{B} \zeta^{B}  \\
& = \frac{1}{2} \left( \ddot{A}_{ij} - \frac{1}{2} A^{km} \dot{A}_{ki} \dot{A}_{mj}\right)  E^{i}{}_A   E^{j}{}_{B} \zeta^{B} \,.
\end{align}
Using the solution $\zeta^A = \bar{B}^A$ and the parallel transport condition (\ref{PT}), one easily shows that
\be
\ddot{\bar{B}}^A =  b_k \ddot{E}^A{}_i H^{ki}\,.
\ee
Plugging this in the previous equation, it reduces to
\be
\label{TPT}
\delta_{AB} \ddot{E}^B{}_m = \frac{1}{2} \left( \ddot{A}_{im} - \frac{1}{2} A^{kn} \dot{A}_{ki} \dot{A}_{nm}\right)  E^{i}{}_A  
\ee
which is nothing else than the time derivative of the parallel transport condition (\ref{PT}). Therefore, the Carrolian boosts $(\bar{B}_{+}, \bar{B}_{-})$ encode the non-trivial solutions to the GDE corresponding to the transverse memory effects in the presence of a non-vanishing initial velocity between the two test particles. Indeed, the solution (\ref{CBGDE}) reproduces the first contribution in Eq.~(\ref{XFNC}). Thus, the choice of symmetry parameter $b_k$ for the boosts corresponds to the choice of momenta $p_i$ in (\ref{XFNC}).

The very same exercise can be reproduced using instead the translational Killing vectors given by $P^{\alpha} \partial_{\alpha} = \chi^i \partial_i $ where $\chi^i = (\chi^x, \chi^y)$ is a constant vector. Evaluated on the reference geodesic, it reads
\be
\bar{P}^{I} \partial_I = E^A{}_{i} (\tau) \chi^i   \partial_A
\ee
such that $\ddot{\bar{P}}^{A} = \ddot{E}^A{}_{i} \chi^i$. Plugging this into the GDE, one finds again that the equation reduces to Eq.~(\ref{TPT}). One recognizes in this solution the second term in (\ref{XFNC}). The choice of parameter $\chi^i$ thus corresponds to a choice of charge $\B^i$ in (\ref{XFNC}). 

Therefore, the Killing symmetries of the geometry naturally carry the relevant information on the different contributions to the memory effects in the transverse direction. In particular, the gauge parameters $(b_i, \chi^i)$ can be related to the different parameters $p^{Y}_i, \B^i_Y$ (i.e. the initial conditions) which set up the initial configuration of the relative motion between the two test particles.

\subsection{Killing tensors and the longitudinal rescaling }

Let us now discuss how to the trivial solutions $\{ \bar{u}, \tau \bar{u}\}$ to the GDE can be related to the hidden symmetries of the wave. The first trivial solution can be constructed using the CZS theorem stated above by considering the metric as a trivial Killing tensor. Then, the GDE admits the trivial solution given by $N_{\alpha} = \bar{g}_{\alpha \beta} \bar{u}^{\beta} = \bar{u}_{\alpha}$. 

The second trivial solution can be related to the non-trivial Killing tensor (\ref{KTBJRR}). At the level of the geodesic motion, this KT encodes the conformal extension of the symmetry group, and it is thus natural that it provides the object allowing to construct the trivial solution to the GDE associated to the freedom to rescale the affine parameter of the reference geodesic. To see this, we first evaluate the KT on the reference geodesic, which reduces in the Fermi coordinates to
\begin{align}
\bar{K}_{IJ} \rd X^I \rd X^J = - U \left[ 2 \rd U \rd V  + \delta_{AB} \rd X^A \rd X^B\right]\,.
\end{align}
The solution to the GDE is given by
\be
N_{I}\rd X^I = \bar{K}_{IJ} \bar{u}^J \rd X^I = - U \rd V \,.
\ee
This is nothing else than the $4$-velocity of the reference geodesic written in Fermi coordinates rescaled by a factor $-U$. Indeed, transforming this co-vector in the BJR coordinates, one finds
\begin{align}
N_{\alpha} \rd x^{\alpha} = \frac{\partial X^I }{\partial x^{\beta}} N_{I} \rd x^{\beta} & =  \frac{\partial V }{\partial v} N_{V} \rd v + \frac{\partial V }{\partial u} N_{V} \rd u  = - u \rd v
\end{align}
where we recognize in the parenthesis the expression (\ref{velo}). Therefore, we have 
\be
N_{\alpha} \rd x^{\alpha} = - \tau \bar{u}_{\alpha} \rd x^{\alpha}
\ee 
where we have used that $U = \tau$. As expected, it corresponds to one of the trivial solution to the geodesic deviation equation labelled in (\ref{affi}) by the parameter $C_1$. This trivial solution corresponds to the freedom to rescale the affine parameter or proper time of the reference geodesic.

\section{Discussion}\label{sec8}

In this work, have shown that, additionally to the well-known velocity memory, a vacuum gravitational plane wave can also induce a displacement memory on a couple of test particles. The main result of this work is the detailed classification of the conditions under which velocity or displacement memory effects are realized. These conditions have been identified by first solving analytically the parallel transport equation (\ref{PT}) for the polarized wave, providing a starting point for the analysis of the geodesic deviation and its general solution given by Eq.~(\ref{e.f4}). 

We have shown that one has to distinguish between pulse and step profiles in order to properly analyze the conditions for the realization of the different memories. We have focused from the start on the configuration where the relative acceleration $\ddot{\zeta}$ asymptotically vanishes. Then, we have introduced a distinction between i) a velocity memory (VM) effect corresponding to a constant shift in the relative velocity, ii) the case where one has no change in the asymptotic relative velocity (VM0) and iii) a constant displacement memory (DM) effect which corresponds to a constant shift in the asymptotic relative distance between the two particles. Notice that in the case of a VM0, one can find configurations where there is no DM but the wave still induce a change in the particles positions, for example by switching their position. This is why one has to distinguish this case from the others.

We have found that pulse profiles generically enjoy a constant VM. The VM0 is realized if the condition (\ref{e.Cpulse0}) is satisfied which corresponds to the following 3 situations:
 \begin{enumerate}
 \item whatever the waves profile if $\dot \A_f-\dot \A_0\not=0$ and if $(\B,p)$ are related by Eq.~(\ref{e.Cpulse1}). This case is illustrated in Fig~\ref{fig:3};
 \item for $p=0$ and all $\B$ if the waves profile is such that $\dot \A_f-\dot \A_0=0$;
 \item for all $(\B,p)$ when the wave profile satisfies the condition~(\ref{e.Cpulse2}).
 \end{enumerate}
 The DM is realized if condition (\ref{condDM}) is satisfied which corresponds in turn to the following three cases:
 \begin{enumerate}
\item If $\dot \A_f \not=0$, then a pure DM requires $p$ and $\B$ to be related by Eq.~(\ref{condDMm}).
Since $\B\not=0$, the DM is possible if and only if $p\not=0$. This case is illustrated in Fig~\ref{fig:3}.
\item If $\dot \A_f=0$ then, whatever the wave profile and whatever $\B$, one has a DM if $p=0$. If $p\neq 0$, then a DM occurs if and only if Eq.~(\ref{e.Cpulse2}) is satisfied.
 \end{enumerate}
 This set of conditions provides the main result of our work and a first concrete classification for the memories to occur.
 
 For step profiles characterized by the two constants $(\lambda_0, \lambda_f)$, the situation is more subtle. We have focused on the case where $\lambda_0 =0$ while $\lambda_f$  remains free. We point out that contrary to the pulse profiles which have received considerable attention in previous investigations of memory effects in vacuum gravitational plane wave \cite{Zhang:2017rno, Zhang:2017geq, Zhang:2018gzn, Zhang:2018srn, Divakarla:2021xrd, Chakraborty:2022qvv}, step profiles have not been studied in this context. Nevertheless, when modeling the form of the non-oscillating contribution to the strain, the form of the wave-profile is similar to a step profile rather than a pulse profile as discussed in Ref~\cite{Favata:2010zu, Favata:2011qi, Hait:2022ukn}. See for example Eqs.~(3.11) and~(3.13) in Ref.~\cite{Favata:2011qi} and the following discussion. In general, one component of the geodesic deviation vector diverges at late time while the second oscillates. This is depicted in Fig~\ref{fig:4}. Therefore, one first has to select the subset of situations where this divergence is cancelled, which corresponds to imposing condition (\ref{Naccf}). Then, within this class of waves, we have shown that the asymptotic value of the relative velocity (\ref{velfut}) generically vanishes independently of the value of the relative velocity in the asymptotic past (\ref{velpast}). It follows that provided (\ref{Naccf}) holds, 
  \begin{enumerate}
 \item A VM occurs generically for a step profile independently of the initial condition $(\zeta_{-}, \dot{\zeta}_{-})$. It corresponds to a cancellation of the initial relative velocity. This case is illustrated in Fig~\ref{fig:6};
 \item A VM0 occurs if and only if condition (\ref{e.ttt}) and (\ref{e.CstepVM0cond}) are satisfied;
 \item A DM occurs if and only if condition (\ref{e.ttt}) and (\ref{e.CstepVM0cond}) and additionally if (\ref{DMM}) hold.
 \end{enumerate}
This second set of conditions completes our classification of the different conditions in the case of the step profiles.

Additionally to the derivation of the analytic conditions for the realization of the different memories, we have provided explicit examples of wave profiles satisfying the different conditions, illustrating the consistency  of our classification. In the end, our results provide a clean understanding, both analytical and numerical, of the displacement and velocity memory effects in a vacuum gravitational plane wave. In particular, it allows one to understand the recent finding presented in Ref.~\cite{Zhang:2024uyp} where the authors have reported two examples of (pulse) wave-profiles inducing a displacement memory. The "magic numbers" responsible for this effect can be understood in terms of the analytical conditions presented for the first time in this work. The classification is also extended in appendix~\ref{Long} to the different memories in the longitudinal direction.
 
 To finish, we have presented two additional results on the physics of vacuum gravitational plane waves. First, we have shown that the vacuum Einstein equations~(\ref{fieldeq}) written in BJR coordinates can be reformulated as a balance equation between the Schwarzian derivatives of the two components of the wave profile (described by auxiliary fields) given by Eq.~(\ref{sch}). This shows that the wave profile enjoys a so far unoticed M\"{o}bius invariance which allows one to generate new profiles from known solutions. This is to be compared to the recent symmetry discussed in Ref.~\cite{Zhao:2024xzo}. Second, we have revisited the problem of solving the geodesic deviation equation in terms of the underlying symmetries using the CZS theorem \cite{Caviglia0}. We have shown how the different trivial and non-trivial solutions of the GDE, relevant for the memories, can be constructed using the generators of the isometries as well as the hidden symmetries. This provides a clean picture of the equivalence between the geodesic deviation equation and the geodesic equation in a vacuum gravitational plane wave from the point of view of its underlying symmetries. This approach can be extended to more complicated radiative spacetimes to analyze the memories, a task which we plan to work out in the near future. Finally, it would be interesting to apply this strategy to analyze the memory effects beyond general relativity. See \cite{Tahura:2020vsa, Tahura:2021hbk, Godazgar:2022pbx, Seraj:2021qja, BenAchour:2024zzk, BenAchour:2024tqt, Heisenberg:2023prj, Siddhant:2020gkn} for related investigations.

\newpage
\appendix

\section{Christoffels and curvature}\label{appA}

The matricial form of the metric is given by
\begin{align}
g_{\mu\nu} = \mat{ccc}{0& 1 & 0 \\1& 0 & 0 \\ 0 & 0& A_{ij}} \qquad g^{\mu\nu} = \mat{ccc}{ 0& 1 & 0 \\1& 0 & 0  \\ 0 & 0 & A^{ij} }
\end{align}
such that the inverse metric reads
\be
\rd s^2 = 2 \partial_u \partial_v + A^{ij} \partial_i \partial_j  \,.
\ee
The non vanishing Christoffel symbols are not modified and are given by
\begin{align}
& \Gamma^{v}{}_{ij} = - \frac{1}{2} \dot{A}_{ij} \;\qquad  \Gamma^i{}_{uj} = \frac{1}{2} A^{i\ell} \dot{A}_{j\ell}\,.
\end{align}
These formulas are coherent with mathematica.
Now, let us recall that
 \be
 \dot{A}^{ik}A_{kj} = -  A^{ik} \dot{A}_{kj}   \qquad \dot{A}_{\ell j} =  - A_{i\ell}A_{kj}   \dot{A}^{ik}\,.
 \ee
 We can compute the Riemann and Ricci tensors. They are given by
 \begin{align}
 R^{\alpha}{}_{\beta\gamma\delta} = \partial_{\gamma} \Gamma^{\alpha}{}_{\beta\delta} - \partial_{\delta} \Gamma^{\alpha}{}_{\beta\gamma} + \Gamma^{\alpha}{}_{\mu\gamma} \Gamma^{\mu}{}_{\beta\delta} - \Gamma^{\alpha}{}_{\mu\delta} \Gamma^{\mu}{}_{\beta\gamma} \,.
 \end{align}
 Since one has $\Gamma^u{}_{\beta\gamma} = 0$ and $\Gamma^{\alpha}{}_{v\gamma} = 0$, we have that $R^{u}{}_{\beta\gamma\delta} = 0$ and $R^{\alpha}{}_{\beta\gamma v} =0$. Now, we list the non-vanishing components which will be useful for the geodesic deviation equation. 

First, we find that the only non-vanishing component of $R^i{}_{\beta\gamma j}$ and $R^i{}_{\beta\gamma u}$  are
 \begin{align}
 R^i{}_{uuj}  =- R^i{}_{uju} =  \dot{\Gamma}^i{}_{uj} + \Gamma^i{}_{ku} \Gamma^k{}_{uj}  - \Gamma^i{}_{\mu j} \Gamma^{\mu}{}_{uu} 
 & = \frac{1}{2} \partial_u (A^{i\ell} \dot{A}_{\ell j}) + \frac{1}{4} A^{i\ell} A^{km} \dot{A}_{k\ell} \dot{A}_{mj}  \\
 & = \frac{1}{2} A^{i\ell} \ddot{A}_{\ell j}  + \frac{1}{2} \dot{A}^{i\ell} \dot{A}_{\ell j} +\frac{1}{4} A^{i\ell} A^{km} \dot{A}_{k\ell} \dot{A}_{mj}  \\
 & = \frac{1}{2} A^{i\ell} \ddot{A}_{j\ell} - \frac{1}{4} A^{i\ell} A^{km} \dot{A}_{k\ell} \dot{A}_{mj}  \\
 & =- R^i{}_{uju}
 \end{align}
 while one has
 \begin{align}
 R^v{}_{jui} & = \partial_u \Gamma^v{}_{ij} - \Gamma^v{}_{ki}\Gamma^k{}_{jk}  = - \frac{1}{2} \left[ \ddot{A}_{ij} - \frac{1}{2} A^{k\ell} \dot{A}_{ki} \dot{A}_{j\ell}\right]\,.
 \end{align}
 This gives us
 \begin{align}
 R_{ujui} = g_{uv} R^v{}_{jui} = - \frac{1}{2} \left[ \ddot{A}_{ij} - \frac{1}{2} A^{k\ell} \dot{A}_{ki} \dot{A}_{j\ell}\right],, \qquad R_{iuuj} = g_{im}  R^m{}_{uuj} = \frac{1}{2} \left( \ddot{A}_{ij} - \frac{1}{2} A^{k\ell} \dot{A}_{ki} \dot{A}_{nj} \right)\,.
 \end{align}
 It follows that the only non-vanishing component of the Ricci tensor reads
 \begin{align}
 R_{\beta\delta} = g^{\mu\gamma} R_{\mu\beta\gamma\delta}\,, \qquad R_{uu} = g^{\mu\gamma} R_{\mu uu \gamma} = A^{ij} R_{iuuj}  =  \frac{1}{2} \left( A^{ij} \ddot{A}_{ij} - \frac{1}{2}A^{ij} A^{k\ell} \dot{A}_{ki} \dot{A}_{nj} \right)\,.
 \end{align}
 and one has obviously that $R=0$.
 
 \section{Wave profile in different coordinates: From $A_{ij}$ to $H_{AB}$}\label{appB}

From Eq.~(\ref{FERMIPROF}), we have
\be
H_{AB}=\frac{1}{2}\left[ \ddot A_{il} -\frac{1}{2}\dot A_{ij} A^{jk}\dot A_{k\ell} \right] E^i_A E^\ell_B \equiv h_{i\ell} E^i_A E^\ell_B
\ee
where $h_{ij}$ is actually related to the Riemann tensor by $h_{ij}=R_{iuuj}$. Let us try to investigate the relations between $A_{ij}$ and $h_{ij}$.

\begin{itemize}
\item First, consider the situations in which $A_{ij}$ depends only on one free function
\be
A_{ij}={\cal A}^2\sigma_{ij}
\ee
where we have introduced the notation $\sigma_{ij}=\{ \delta_{ij},\varepsilon_{ij},\tau_{ij}\}$ for the following 2 $\times$ 2 matrices
\begin{equation}
\delta_{ij}=\left(\begin{array}{ll} 1 &0 \\ 0 & 1 \end{array}  \right), \quad
\varepsilon_{ij} = \left(\begin{array}{ll} 0 &1 \\ 1 & 0 \end{array}  \right), \quad
\tau_{ij} = \left(\begin{array}{ll} 1 &0 \\ 0 & -1 \end{array}  \right).
%\omega_{ij} = \left(\begin{array}{ll} 0 &-1 \\ 1 & 0 \end{array}  \right).
\end{equation}
One deduces that $R_{uu}=2 \ddot {\cal A}/{\cal A}=\ddot A/A -(\dot A/A)^2/2$ so that the Einstein equations imply that
\be
\ddot {\cal A} = 0.
\ee
Since $h_{ij}= {\cal A}\ddot{\cal A}\sigma_{ij}$, we deduce that for all solutions of the vacuum
\be
H_{AB}=0\,.
\ee 
\item Then, consider the case with 2 free functions. First if
$$
A_{ij} =  \mat{cc}{{\cal A}_1^2&  0 \\ 0 & {\cal A}^2_2 } ,
$$
which is the case studied in the core of this article and generally in the literature, then $R_{uu}=0$ gives
$$
\frac{\ddot {\cal A}_1}{{\cal A}_1} + \frac{\ddot {\cal A}_2}{{\cal A}_2} =0
$$
while
\be
h_{ij} =  \mat{cc}{{\cal A}_1\ddot {\cal A}_1&  0 \\ 0 & {\cal A}_2\ddot {\cal A}_2 } 
\ee
which is indeed traceless since $h_{ij}A^{ij}=0$ from the Einstein equations. Then, the equation of evolution of $E^i_A$ is
\be
\dot E^i_A = -\frac{\dot{\cal A}_i}{{\cal A}_i}  E^i_A
\ee
without summation on $i$, 
since
$$
-\frac{1}{2}A^{ij}\dot A_{ij}= - \mat{cc}{ \frac{\dot{\cal A}_1}{{\cal A}_1} & 0 \\ 0& \frac{\dot{\cal A}_2}{{\cal A}_2}}
$$
which integrates easily as
\be
E^i_A(u) = \frac{{\cal A}_i(u_0)}{{\cal A}_i(u)} E^i_A(u_0)\,.
\ee
It follows that $H_{AB}$ is of the form
\be
H_{AB} = \mat{cc}{H_+& 0 \\ 0 & -H_+}
\ee 
with $H_+=\ddot{\cal A}_1/{\cal A}_1$. This is the case we already studied.
\item Now let us show that the $\times$ polarisation can be obtained from the decomposition
\begin{equation}
A_{ij} =  \mat{cc}{{\cal A}^2&   {\cal A}_{12}^2 \\  {\cal A}_{12}^2 & {\cal A}^2 } 
\end{equation}
and let us use the redefinitions
\be
{\cal A}^2=\frac{1}{2}\left( {\cal A}^2_{11}+{\cal A}^2_{22} \right),
\quad
{\cal A}_{12}^2= \frac{1}{2}\left( {\cal A}^2_{22}-{\cal A}^2_{11} \right)\,.
\ee
First, thanks to the rotation defined by
\be
{R^\ell}_j  = \frac{1}{\sqrt{2}} \mat{cc}{1&  -1 \\ 1 &1 } 
\ee
we have that
\be
A_{ij} = {R_i}^k    \tilde A_{k\ell} {R^\ell}_j 
\quad\hbox{with}\quad
 \tilde A_{k\ell}   =  \mat{cc}{{\cal A}_{11}^2&  0 \\ 0 &{\cal A}_{22}^2} .
\ee
Since ${R_i}^k $ is constant and satisfies ${R_i}^k{R^i}_\ell = \delta^k_\ell$ (that is $^{t}RR=R^{-1}R=Id$). With these notations we have $A=R^{-1} \tilde A R$ and $A^{-1}=R^{-1} A^{-1} R$. First, we get that $R_{uu}[A]=R_{uu}[\tilde A]$ so that the vacuum Einstein equations reduce to
$$
\frac{\ddot {\cal A}_1}{{\cal A}_1} + \frac{\ddot {\cal A}_2}{{\cal A}_2} =0\,.
$$
Then, the evolution of the dyad, $\dot E^i_A = -\frac{1}{2}A^{i k}\dot A_{k j} E^j_A = -\frac{1}{2}A^{ij}\dot A_{ij} E^j_A $,simplifies to  
$$
\frac{\dd}{\dd u}({R^\ell}_i  E^i_A) = -\frac{1}{2}\tilde A^{\ell k} \dot{\tilde A}_{k p}   ( {R^p}_j E^j_A )\,.
$$
Hence the vector base after rotation, $\tilde E^i_A\equiv {R^\ell}_i  E^i_A$, it becomes
\be
\dot{ \tilde E}^i_A = -\frac{\dot{\cal A}_i}{{\cal A}_i}  \tilde E^i_A
\ee
without summation on $i$, 
since
$$
-\frac{1}{2}\tilde A^{ij}\dot {\tilde A}_{ij}= - \mat{cc}{ \frac{\dot{\cal A}_1}{{\cal A}_1} & 0 \\ 0& \frac{\dot{\cal A}_2}{{\cal A}_2}}
$$
which integrates easily as
\be
\tilde E^i_A(u) = \frac{{\cal A}_i(u_0)}{{\cal A}_i(u)}\tilde E^i_A(u_0).
\ee
To finish
\be
 H_{AB} =  h_{ij}E^i_AE^j_B = {R_k}^i \tilde h_{k\ell} {R^\ell}_j E^i_AE^j_B
 \quad\hbox{with}\quad
 \tilde h_{k\ell} =  \mat{cc}{{\cal A}_1\ddot {\cal A}_1&  0 \\ 0 & {\cal A}_2\ddot {\cal A}_2 } \,.
\ee
\end{itemize}

\section{Apparent divergences in the algebraic solution~(\ref{e.f4}) of the equation of motion}\label{appC}
 
 In this appendix, we discuss why the divergence in Eq.~(\ref{e.f4}) is only an apparent divergence without physical meaning.  As can be seen from Eq.~(\ref{e.f4}) the displacement,
\be
 \zeta^i(u) = {\cal A}_i(u)\left[ H^{ii}(u_0,u) p_i - \B^i \right] \nonumber
\ee
may seem singular when ${\cal A}_{1}$ or ${\cal A}_{2}$ changes sign. In particular because of Eq.~(\ref{wavBrink1}) one concludes that it shall happen at least for one of the two modes.  The term $H^{ii}(u_0,u)$ may diverge when ${\cal A}_i$ vanishes. But this divergence shall be compensated in a way that ${\cal A}_i H^{ii}(u_0,u)$ remains regular since indeed the differential equation $(\ddot \zeta_1,\ddot \zeta_2)=(H_+\zeta_1,-H_+\zeta_2)$ is well-behaved as long as $H_+$ is non-singular.
 
To show the singularity of the analytic solution is only apparent, let us consider the function
 $$
  {\cal A}(u)\int_{u_0}^u\frac{\dd v}{{\cal A}^2(v)}
 $$
 in a neighborhood of $u_*>u_0$ where ${\cal A}(u_*)=0$. Expanding locally ${\cal A}$ around $u_*$ as
 $$
 {\cal A}(u)=\dot{\cal A}_*(u-u_*) 
 $$
 with $\dot{\cal A}_*\not=0$, which shall be the case since otherwise the differential equation~(\ref{wavBrink1}) for ${\cal A}$ would imply that ${\cal A}_*,\dot{\cal A}_*$ and $\ddot{\cal A}_*$ vanish so that ${\cal A}=0$ for all $u$, which we exclude. It follows that
 \begin{align}
 {\cal A}(u)\int_{u_0}^u\frac{\dd v}{{\cal A}^2(v)} &= {\cal A}(u)\left[\int_{u_0}^{u_*-\epsilon}\frac{\dd v}{{\cal A}^2(v)} 
 +  \int_{u_* - \epsilon}^{u}\frac{\dd v}{{\cal A}^2(v)} \right] \nonumber \\
 &= \dot{\cal A}_*(u-u_*) \int_{u_0}^{u_*-\epsilon}\frac{\dd v}{{\cal A}^2}  + \dot{\cal A}_*(u-u_*) \int_{u_* - \epsilon}^{u}\frac{\dd v}{\dot{\cal A}_*^2(v-u_*)^2} \nonumber\\
 &=\dot{\cal A}_*(u-u_*) \int_{u_0}^{u_*-\epsilon}\frac{\dd v}{{\cal A}^2}  -\frac{1}{\dot{\cal A}_*} (u-u_*) \left[ \frac{1}{\epsilon} + \frac{1}{u-u_*} \right]\nonumber\\
 &=\dot{\cal A}_*(u-u_*) \int_{u_0}^{u_*-\epsilon}\frac{\dd v}{{\cal A}^2}  -\frac{1}{\dot{\cal A}_*}  \left[ \frac{(u-u_*)}{\epsilon} + 1 \right]\,.
 \end{align}
 Hence
 \be
  {\cal A}(u_*)\int_{u_0}^{u_*}\frac{\dd v}{{\cal A}^2(v)} = -\frac{1}{\dot {\cal A}_*}\,,
 \ee
 which is finite even though $\int_{u_0}^{u_*}\frac{\dd v}{{\cal A}^2(v)}$ is infinite and ${\cal A}(u_*)=0$.

\section{Integrability of the geodesic deviation equation}\label{GDE}

In this appendix, we review the proof of the CZS theorem, showing that for a rank-2 Killing tensor, the vector 
\be
N_{\mu} = \bar{K}_{\mu\nu} \bar{u}^{\nu}
\ee
is an exact solution to the first order geodesic deviation equation (\ref{gde1}). Here, a bar means that the quantity is evaluated on the reference geodesic.

To prove this result, we will use the definition of the Killing tensor and the commutation rules of the covariant derivative, i.e. 
\begin{align}
\label{def}
  \nabla_{\mu} K_{\nu\alpha} + \nabla_{\nu} K_{\alpha\mu}  & = - \nabla_{\alpha} K_{\mu\nu} \\
  \label{com}
  [\nabla_{\rho}, \nabla_{\alpha}] K_{\mu\nu} & = +R_{\mu\lambda\alpha\rho} K^{\lambda}{}_{\nu} +R^{\lambda\nu\alpha\rho} K_{\mu}{}^{\lambda} \,.
\end{align}
Notice here that there is a choice in the definition of the Riemann tensor.
Now, taking the covariant derivative along a vector $u^{\rho} \partial_{\rho}$ and projecting along the Killing tensor directions, we have the following equalities
\begin{align}
    u^{\mu} u^{\nu} u^{\rho} \nabla_{\rho} (\nabla_{\mu} K_{\nu\alpha} + \nabla_{\nu} K_{\alpha\mu} )   = \left\{
  \begin{array}{ll}
        2 u^{\mu} u^{\nu} u^{\rho} \nabla_{\rho}  (\nabla_{\mu} K_{\nu\alpha})  &  \\
        -  u^{\mu} u^{\nu}  u^{\rho} ( \nabla_{\rho}  \nabla_{\alpha} K_{\mu\nu} ) & 
    \end{array}
\right.
\end{align}
where the first one is simply using the symmetry in $(\mu\nu)$ and the second is the Killing tensor equation. Starting from this, we write
\begin{align}
  2 u^{\mu} u^{\nu} u^{\rho} \nabla_{\rho}  (\nabla_{\mu} K_{\nu\alpha})   & = -  u^{\mu} u^{\nu}  u^{\rho} ( \nabla_{\rho}  \nabla_{\alpha} K_{\mu\nu} ) \nn \\
  & = -  u^{\mu} u^{\nu}  u^{\rho} ( \nabla_{\alpha} \nabla_{\rho} K_{\mu\nu} + R^{\lambda}{}_{\mu\alpha\rho} K_{\lambda \nu} +R^{\lambda}{}_{\nu\alpha\rho} K_{\mu\lambda} ) \nn \\
  & = - 2R^{\lambda}{}_{\nu\alpha\rho} K_{\lambda\mu} u^{\mu} u^{\nu} u^{\rho} - u^{\mu} u^{\nu} u^{\rho} \nabla_{\alpha} \nabla_{\rho} K_{\mu\nu}\,.
\end{align}
In the second line, we have used the commutation relation of the covariant derivative (\ref{com}). The last line is obtained using again the appropriate symmetry in the indices. The resulting equality is given by
\begin{align}
 2 u^{\mu} u^{\nu} u^{\rho} \nabla_{\rho}  (\nabla_{\mu} K_{\nu\alpha}) + 2R^{\lambda}{}_{\nu\alpha\rho} K_{\lambda\mu} u^{\mu} u^{\nu} u^{\rho}  =  - u^{\mu} u^{\nu} u^{\rho} \nabla_{\alpha} \nabla_{\rho} K_{\mu\nu}\,.
\end{align}
Now, using that $u^{\mu}\partial_{\mu}$ is geodesic, i.e. $u^{\mu} \nabla_{\mu} u^{\rho} =0$, we can write
\begin{align}
u^{\mu}  u^{\rho}  \nabla_{\rho}  \nabla_{\mu} ( K_{\alpha\nu} u^{\nu} ) + R^{\lambda}{}_{\nu\alpha\rho} (K_{\lambda\mu} u^{\mu} ) u^{\nu} u^{\rho}  =  -  \frac{1}{2}u^{\mu} u^{\nu} u^{\rho} \nabla_{\alpha} \nabla_{\rho} K_{\mu\nu}\,.
\end{align}
Finally, introducing the vector $N_{\mu} = K_{\mu\nu} u^{\nu}$ and symmetrizing the last term, we find
\begin{align}
u^{\mu}  u^{\rho} \nabla_{\rho}  \nabla_{\mu} n_{\alpha} + R^{\lambda}{}_{\nu\alpha\rho} n_{\lambda} u^{\nu} u^{\rho}  =  - \frac{1}{6}u^{\mu} u^{\nu} u^{\rho} \nabla_{\alpha} \nabla_{(\rho} K_{\mu\nu)} =0
\end{align}
where we have used in the final step the definition of the Killing tensor, i.e. $ \nabla_{(\rho} K_{\mu\nu)} =0$ . It follows that 
\be
\bar{u}^{\mu}  \bar{u}^{\rho} \nabla_{\rho}  \nabla_{\mu} N_{\alpha}   = \bar{R}_{\alpha\rho\nu\lambda} \bar{u}^{\nu} \bar{u}^{\rho} N^{\lambda} 
\ee
which is the first order geodesic deviation equation.

 \section{Longitudinal memory}\label{Long}

In this appendix, we complete our classification on the conditions for having a displacement or velocity memory effect by analyzing the longitudinal motion.
Let us consider the motion of the test particles in the direction of the propagation of the wave. Using the coordinates
\begin{align}\label{e.defZT}
Z (u) =  \frac{1}{2}U(u) + V(u)\,, \qquad T (u)  =  V(u)  - \frac{1}{2}U(u)\,,
\end{align}
it is clear that the longitudinal motion is determined only by the function $V(u)$ given by Eq.~(\ref{FNC2}). Thanks to the expressions~ (\ref{vtraj}) and~(\ref{xtraj}) for $x^i(u)$ and $v(u)$, one gets
\begin{align}
V(u) & = v_0 + \frac{\epsilon}{2}(u-u_0) +  \frac{1}{4} \dot{A}_{ij} \B^i \B^j +  \frac{1}{4} \left[ \left(\dot{A}_{ij} H^{ik} H^{j\ell} - 2 H^{k\ell}\right) p_{\ell} + \frac{1}{2} \dot{A}_{ij} \B^i H^{jk}\right] p_k\,.
\end{align}
Proceeding as in the previous section, we consider two geodesics. The reference one defined by the constants of motion $\bar{p}_i =0$ and $\bar{\B}^i =0$ admits the following V-trajectory
\be
\bar V(u)  =  \frac{\epsilon}{2}(u-u_0)
\ee
which consists in the standard inertial drift in Minkowski. The initial position and momentum of the second test particle is still labelled by  $(p_i, \B^i)$. The relative longitudinal distance between the two geodesics is given by
\begin{align}
\zeta^Z & = Z(u) - \bar{Z}(u)  =  V(u) - \bar{V}(u) 
\end{align}
which allows one to compute  
\begin{align}
\zeta^Z & = v_0 +  \frac{1}{4} \dot{A}_{ij} \B^i \B^j +  \frac{1}{4} \left[ \left(\dot{A}_{ij} H^{ik} H^{j\ell} - 2 H^{k\ell}\right) p_{\ell} + \frac{1}{2} \dot{A}_{ij} \B^i H^{jk}\right] p_k\\
\dot{\zeta^Z} & = \frac{1}{4} \ddot{A}_{ij} \B^i \B^j +  \frac{1}{4} \left[ \ddot{A}_{ij} \left( H^{ik} H^{j\ell} p_{\ell} + \frac{1}{2} \B^i H^{jk}\right) + \dot{A}_{ij} (A^{ik} H^{j\ell} + A^{j\ell} H^{ik} + \frac{1}{2} A^{jk} \B^i)\right] p_k\,.
\end{align}
It provides the general expressions to analyze the relative longitudinal motion w.r.t. the initial conditions $(p_i, \B^i)$. In the following, we shall be interested in the case where $p_i =0$, which corresponds to the second particle being at rest initially in the 2D transverse plane. Then, the conditions simply reduce to 
\begin{align}
\zeta^Z  = v_0 +  \frac{1}{4} \dot{A}_{ij} \B^i \B^j  \qquad \dot{\zeta}^Z  = \frac{1}{4} \ddot{A}_{ij} \B^i \B^j
\end{align}
From these simple expressions, we obtain 
\begin{itemize}
\item a constant longitudinal velocity memory effect provided
\be
 \dddot{A}_{ij}(u_f) = 0 \qquad  \ddot{A}_{ij}(u_f) \neq 0
\ee
\item a constant longitudinal displacement memory effect if
\be
\ddot{A}_{ii}(u_f) =0 \qquad \dot{A}_{ii} (u_f) \neq 0
\ee
\item no longitudinal memory effects if
\be
\label{condlongno}
\ddot{A}_{ii}(u_f) = 0 \qquad \dot{A}_{ii} (u_f) = 0
\ee
\end{itemize}
This concludes the discussion on memory effects and the associated conditions on the asymptotic form of the wave profile $A_{ii}$.


\newpage 

\begin{thebibliography}{ab}

\bibitem{Zeldovich:1974gvh}
Y.~B.~Zel'dovich and A.~G.~Polnarev,
``Radiation of gravitational waves by a cluster of superdense stars,''
Sov. Astron. \textbf{18} (1974), 17

\bibitem{Christodoulou:1991cr}
D.~Christodoulou,
``Nonlinear nature of gravitation and gravitational wave experiments,''
Phys. Rev. Lett. \textbf{67} (1991), 1486-1489
doi:10.1103/PhysRevLett.67.1486

\bibitem{Blanchet:1992br}
L.~Blanchet and T.~Damour,
``Hereditary effects in gravitational radiation,''
Phys. Rev. D \textbf{46} (1992), 4304-4319
doi:10.1103/PhysRevD.46.4304

\bibitem{Braginsky:1985vlg}
V.~B.~Braginsky and L.~P.~Grishchuk,
``Kinematic Resonance and Memory Effect in Free Mass Gravitational Antennas,''
Sov. Phys. JETP \textbf{62} (1985), 427-430

\bibitem{Bieri:2024ios}
L.~Bieri and A.~Polnarev,
``Gravitational Wave Displacement and Velocity Memory Effects,''
\href{http://arXiv.org/abs/2402.02594}{{\texttt{arXiv:2402.02594}}}.


\bibitem{Pasterski:2015tva}
S.~Pasterski, A.~Strominger and A.~Zhiboedov,
``New Gravitational Memories,''
JHEP \textbf{12} (2016), 053
\href{http://arXiv.org/abs/1502.06120}{{\texttt{arXiv:1502.06120}}}.

\bibitem{Nichols:2018qac}
D.~A.~Nichols,
``Center-of-mass angular momentum and memory effect in asymptotically flat spacetimes,''
Phys. Rev. D \textbf{98} (2018) no.6, 064032
\href{http://arXiv.org/abs/1807.08767}{{\texttt{arXiv:1807.08767}}}.

\bibitem{Seraj:2021rxd}
A.~Seraj and B.~Oblak,
``Gyroscopic gravitational memory,''
JHEP \textbf{11} (2023), 057
\href{http://arXiv.org/abs/2112.04535}{{\texttt{arXiv:2112.04535}}}.


\bibitem{Lasky:2016knh}
P.~D.~Lasky, E.~Thrane, Y.~Levin, J.~Blackman and Y.~Chen,
``Detecting gravitational-wave memory with LIGO: implications of GW150914,''
Phys. Rev. Lett. \textbf{117} (2016) no.6, 061102
\href{http://arXiv.org/abs/1605.01415}{{\texttt{arXiv:1605.01415}}}.

\bibitem{Boersma:2020gxx}
O.~M.~Boersma, D.~A.~Nichols and P.~Schmidt,
``Forecasts for detecting the gravitational-wave memory effect with Advanced LIGO and Virgo,''
Phys. Rev. D \textbf{101} (2020) no.8, 083026
\href{http://arXiv.org/abs/2002.01821}{{\texttt{arXiv:2002.01821}}}.


\bibitem{Hubner:2021amk}
M.~H\"ubner, P.~Lasky and E.~Thrane,
``Memory remains undetected: Updates from the second LIGO/Virgo gravitational-wave transient catalog,''
Phys. Rev. D \textbf{104} (2021) no.2, 023004
\href{http://arXiv.org/abs/2105.02879}{{\texttt{arXiv:2105.02879}}}.

\bibitem{Goncharov:2023woe}
B.~Goncharov, L.~Donnay and J.~Harms,
``Inferring fundamental spacetime symmetries with gravitational-wave memory: from LISA to the Einstein Telescope,''
\href{http://arXiv.org/abs/2310.10718}{{\texttt{arXiv:2310.10718}}}.


\bibitem{Strominger:2014pwa}
A.~Strominger and A.~Zhiboedov,
``Gravitational Memory, BMS Supertranslations and Soft Theorems,''
JHEP \textbf{01} (2016), 086
\href{http://arXiv.org/abs/1411.5745}{{\texttt{arXiv:1411.5745}}}.

\bibitem{Strominger:2017zoo}
A.~Strominger,
``Lectures on the Infrared Structure of Gravity and Gauge Theory,''
\href{http://arXiv.org/abs/1703.05448}{{\texttt{arXiv:1703.05448}}}.

\bibitem{Barnich:2009se}
G.~Barnich and C.~Troessaert,
``Symmetries of asymptotically flat 4 dimensional spacetimes at null infinity revisited,''
Phys. Rev. Lett. \textbf{105} (2010), 111103
\href{http://arXiv.org/abs/0909.2617}{{\texttt{arXiv:0909.2617}}}.

\bibitem{Campiglia:2014yka}
M.~Campiglia and A.~Laddha,
``Asymptotic symmetries and subleading soft graviton theorem,''
Phys. Rev. D \textbf{90} (2014) no.12, 124028
\href{http://arXiv.org/abs/1408.2228}{{\texttt{arXiv:1408.2228}}}.

\bibitem{Campiglia:2016efb}
M.~Campiglia and A.~Laddha,
``Sub-subleading soft gravitons and large diffeomorphisms,''
JHEP \textbf{01} (2017), 036
\href{http://arXiv.org/abs/1608.00685}{{\texttt{arXiv:1608.00685}}}.

\bibitem{Compere:2019odm}
G.~Comp\`ere,
``Infinite towers of supertranslation and superrotation memories,''
Phys. Rev. Lett. \textbf{123} (2019) no.2, 021101
\href{http://arXiv.org/abs/1904.00280}{{\texttt{arXiv:1904.00280}}}.

\bibitem{Strominger:2021mtt}
A.~Strominger,
``$w_{1+\infty}$ Algebra and the Celestial Sphere: Infinite Towers of Soft Graviton, Photon, and Gluon Symmetries,''
Phys. Rev. Lett. \textbf{127} (2021) no.22, 221601
\href{http://arXiv.org/abs/2105.14346}{{\texttt{arXiv:2105.14346}}}.

\bibitem{Freidel:2021fxf}
L.~Freidel, R.~Oliveri, D.~Pranzetti and S.~Speziale,
``The Weyl BMS group and Einstein\textquoteright{}s equations,''
JHEP \textbf{07} (2021), 170
\href{http://arXiv.org/abs/2104.05793}{{\texttt{arXiv:2104.05793}}}.

\bibitem{Freidel:2021dfs}
L.~Freidel, D.~Pranzetti and A.~M.~Raclariu,
``Sub-subleading soft graviton theorem from asymptotic Einstein\textquoteright{}s equations,''
JHEP \textbf{05} (2022), 186
\href{http://arXiv.org/abs/2111.15607}{{\texttt{arXiv:2111.15607}}}.

\bibitem{Blanchet:2023pce}
L.~Blanchet, G.~Comp\`ere, G.~Faye, R.~Oliveri and A.~Seraj,
``Multipole expansion of gravitational waves: memory effects and Bondi aspects,''
JHEP \textbf{07} (2023), 123
\href{http://arXiv.org/abs/2303.07732}{{\texttt{arXiv:2303.07732}}}.

\bibitem{Geiller:2024bgf}
M.~Geiller,
``Celestial $w_{1+\infty}$ charges and the subleading structure of asymptotically-flat spacetimes,''
\href{http://arXiv.org/abs/2403.05195}{{\texttt{arXiv:2403.05195}}}.

\bibitem{Hawking:2016msc}
S.~W.~Hawking, M.~J.~Perry and A.~Strominger,
``Soft Hair on Black Holes,''
Phys. Rev. Lett. \textbf{116} (2016) no.23, 231301
\href{http://arXiv.org/abs/1601.00921}{{\texttt{arXiv:1601.00921}}}.

\bibitem{Donnay:2018ckb}
L.~Donnay, G.~Giribet, H.~A.~Gonz\'alez and A.~Puhm,
``Black hole memory effect,''
Phys. Rev. D \textbf{98} (2018) no.12, 124016
\href{http://arXiv.org/abs/1809.07266}{{\texttt{arXiv:1809.07266}}}.

\bibitem{Bhattacharjee:2020vfb}
S.~Bhattacharjee, S.~Kumar and A.~Bhattacharyya,
``Displacement memory effect near the horizon of black holes,''
JHEP \textbf{03} (2021), 134
\href{http://arXiv.org/abs/2010.16086}{{\texttt{arXiv:2010.16086}}}.

\bibitem{Sarkar:2021djs}
S.~Sarkar, S.~Kumar and S.~Bhattacharjee,
``Can we detect a supertranslated black hole?,''
Phys. Rev. D \textbf{105} (2022) no.8, 084001
\href{http://arXiv.org/abs/:2110.03547}{{\texttt{arXiv::2110.03547}}}.


\bibitem{Flanagan:2018yzh}
\'E.~\'E.~Flanagan, A.~M.~Grant, A.~I.~Harte and D.~A.~Nichols,
``Persistent gravitational wave observables: general framework,''
Phys. Rev. D \textbf{99} (2019) no.8, 084044
\href{http://arXiv.org/abs/1901.00021}{{\texttt{arXiv:1901.00021}}}.


\bibitem{Grant:2021hga}
A.~M.~Grant and D.~A.~Nichols,
``Persistent gravitational wave observables: Curve deviation in asymptotically flat spacetimes,''
Phys. Rev. D \textbf{105} (2022) no.2, 024056
[erratum: Phys. Rev. D \textbf{107} (2023) no.10, 109902]
\href{http://arXiv.org/abs/2109.03832}{{\texttt{arXiv:2109.03832}}}.

\bibitem{Seraj:2022qqj}
A.~Seraj and T.~Neogi,
``Memory effects from holonomies,''
Phys. Rev. D \textbf{107} (2023) no.10, 10
\href{http://arXiv.org/abs/2206.14110}{{\texttt{arXiv:2206.14110}}}.

\bibitem{Grant:2023ged}
A.~M.~Grant,
``Persistent gravitational wave observables: Nonlinearities in (non-)geodesic deviation,''
\href{http://arXiv.org/abs/2401.00047}{{\texttt{arXiv:2401.00047}}}.



\bibitem{Siddhant:2024nft}
S.~Siddhant, A.~M.~Grant and D.~A.~Nichols,
``Higher memory effects and the post-Newtonian calculation of their gravitational-wave signals,''
\href{http://arXiv.org/abs/2403.13907}{{\texttt{arXiv:2403.13907}}}.

\bibitem{Compere:2018ylh}
G.~Comp\`ere, A.~Fiorucci and R.~Ruzziconi,
``Superboost transitions, refraction memory and super-Lorentz charge algebra,''
[erratum: JHEP \textbf{04} (2020), 172]
\href{http://arXiv.org/abs/1810.00377}{{\texttt{arXiv:1810.00377}}}.


\bibitem{Roche:2022bcz}
C.~Roche, A.~B.~Aazami and C.~Cederbaum,
``Exact parallel waves in general relativity,''
Gen. Rel. Grav. \textbf{55}, no.2, 40 (2023)
\href{http://arXiv.org/abs/2207.03591}{{\texttt{arXiv:2207.03591}}}.

\bibitem{PenroseLimit}
R.~Penrose, ``Any space-time has a plane wave as a limit'', in Differential Geometry and Relativity: A Volume in Honour of Andr\'e Lichnerowicz on His 60th Birthday, edited by M. Cahen and M. Flato (Springer Netherlands,
Dordrecht, 1976), pp. 271–275.

\bibitem{Impulsive}
R. Penrose. The geometry of impulsive gravitational waves. In General relativity (papers in
honour of J. L. Synge), pages 101–115. Clarendon Press, Oxford, 1972.

\bibitem{Steinbauer:1997dw}
R.~Steinbauer,
``Geodesics and geodesic deviation for impulsive gravitational waves,''
J. Math. Phys. \textbf{39}, 2201-2212 (1998)
\href{http://arXiv.org/abs/9710119}{{\texttt{arXiv:9710119}}}.


\bibitem{Zhang:2017jma}
P.~M.~Zhang, C.~Duval and P.~A.~Horvathy,
``Memory Effect for Impulsive Gravitational Waves,''
Class. Quant. Grav. \textbf{35}, no.6, 065011 (2018)
\href{http://arXiv.org/abs/1709.02299}{{\texttt{arXiv:1709.02299}}}.

\bibitem{Steinbauer:2018iis}
R.~Steinbauer,
``The memory effect in impulsive plane waves: comments, corrections, clarifications,''
Class. Quant. Grav. \textbf{36} (2019) no.9, 098001
\href{http://arXiv.org/abs/1811.10940}{{\texttt{arXiv:1811.10940}}}.

\bibitem{Shore:2018kmt}
G.~M.~Shore,
``Memory, Penrose Limits and the Geometry of Gravitational Shockwaves and Gyratons,''
JHEP \textbf{12}, 133 (2018)
\href{http://arXiv.org/abs/1811.08827}{{\texttt{arXiv:1811.08827}}}.

\bibitem{Adamo:2022rob}
T.~Adamo, A.~Cristofoli and P.~Tourkine,
``The ultrarelativistic limit of Kerr,''
JHEP \textbf{02} (2023), 107
\href{http://arXiv.org/abs/2209.05730}{{\texttt{arXiv:2209.05730}}}.

\bibitem{He:2023qha}
T.~He, A.~M.~Raclariu and K.~M.~Zurek,
``From shockwaves to the gravitational memory effect,''
JHEP \textbf{01} (2024), 006
\href{http://arXiv.org/abs/2305.14411}{{\texttt{arXiv:2305.14411}}}.

%%%%%%%%%%%%%%%%%%%% MEMORIES IN PP WAVE


\bibitem{Zhang:2017rno}
P.~M.~Zhang, C.~Duval, G.~W.~Gibbons and P.~A.~Horvathy,
``The Memory Effect for Plane Gravitational Waves,''
Phys. Lett. B \textbf{772}, 743-746 (2017)
\href{http://arXiv.org/abs/1704.05997}{{\texttt{arXiv:1704.05997}}}.

\bibitem{Zhang:2017geq}
P.~M.~Zhang, C.~Duval, G.~W.~Gibbons and P.~A.~Horvathy,
``Soft gravitons and the memory effect for plane gravitational waves,''
Phys. Rev. D \textbf{96}, no.6, 064013 (2017)
\href{http://arXiv.org/abs/1705.01378}{{\texttt{arXiv:1705.01378}}}.


\bibitem{Zhang:2018gzn} 
P.~M.~Zhang, M.~Elbistan, G.~W.~Gibbons and P.~A.~Horvathy,
``Sturm\textendash{}Liouville and Carroll: at the heart of the memory effect,''
Gen. Rel. Grav. \textbf{50}, no.9, 107 (2018)
\href{http://arXiv.org/abs/1803.09640}{{\texttt{arXiv:1803.09640}}}.


\bibitem{Zhang:2018srn}
P.~M.~Zhang, C.~Duval, G.~W.~Gibbons and P.~A.~Horvathy,
``Velocity Memory Effect for Polarized Gravitational Waves,''
JCAP \textbf{05}, 030 (2018)
\href{http://arXiv.org/abs/1802.09061}{{\texttt{arXiv:1802.09061}}}.

\bibitem{Divakarla:2021xrd}
A.~K.~Divakarla and B.~F.~Whiting,
``First-order velocity memory effect from compact binary coalescing sources,''
Phys. Rev. D \textbf{104}, no.6, 064001 (2021)
\href{http://arXiv.org/abs/2106.05163}{{\texttt{arXiv:2106.05163}}}.


\bibitem{Chakraborty:2022qvv}
I.~Chakraborty and S.~Kar,
``A simple analytic example of the gravitational wave memory effect,''
Eur. Phys. J. Plus \textbf{137}, no.4, 418 (2022)
\href{http://arXiv.org/abs/2202.10661}{{\texttt{arXiv:2202.10661}}}.

\bibitem{Elbistan:2023qbp}
M.~Elbistan, P.~M.~Zhang and P.~A.~Horvathy,
``Memory effect \& Carroll symmetry, 50 years later,''
Annals Phys. \textbf{459} (2023), 169535
doi:10.1016/j.aop.2023.169535

\bibitem{Chakraborty:2020uui}
I.~Chakraborty and S.~Kar,
``Memory effects in Kundt wave spacetimes,''
Phys. Lett. B \textbf{808} (2020), 135611
\href{http://arXiv.org/abs/2005.00245}{{\texttt{arXiv:2005.00245}}}.



\bibitem{Zhang:2024uyp}
P.~M.~Zhang and P.~A.~Horvathy,
``Displacement within velocity effect in gravitational wave memory,''
\href{http://arXiv.org/abs/2405.12928}{{\texttt{arXiv:2405.12928}}}.



%%%%%%%%%%%%%%% GDE AND SYMMETRIES

\bibitem{Caviglia0}
G. Caviglia, C. Zordan, and F. Salmistraro, “Equation of geodesic deviation and Killing tensors,” Int. J. Theor. Phys. 21, 391–6 (1982).

\bibitem{Caviglia1}
G. Caviglia, “Dynamical symmetries: an approach to Jacobi fields and to constants of geodesic motion,” J. Math. Phys. 24, 2065–2069 (1983).


\bibitem{Salmistraro:1983xr}
F.~Salmistraro,``A NOTE ON Killing TENSORS IN VACUUM SPACE-TIMES,''
Lett. Nuovo Cim. \textbf{36}, 35-38 (1983)
%doi:10.1007/BF02754908

\bibitem{Dolan}
P. Dolan and N. S. Swaminarayan, “Solutions of the geodesic deviation equation obtained by using hidden symmetries,” Proc. R. Ir. Acad. A: Math. Phys. Sc. 84A, 133–139 (1984).

\bibitem{Cariglia:2018erv}
M.~Cariglia, T.~Houri, P.~Krtous and D.~Kubiznak,
``On Integrability of the Geodesic Deviation Equation,''
Eur. Phys. J. C \textbf{78}, no.8, 661 (2018)
\href{http://arXiv.org/abs/1805.07677}{{\texttt{arXiv:1805.07677}}}.


%%%%%%%%%%%%%%%%%%%%%%%%%%%%%%%%%%%%%%%%%%%%%%

\bibitem{Souriau}
J-M. Souriau, “Ondes et radiations gravitationnelles,” Colloques Internationaux du CNRS No
220, pp. 243-256. Paris (1973).


%%%%%%%%%%%%%%%%%%%%%% SYMMETRIES OF PP WAVE
\bibitem{Sippel:1986if}
R.~Sippel and H.~Gonner,
``Symmetry Classes of $P P$ Waves,''
Gen. Rel. Grav. \textbf{18}, 1229-1243 (1986)
doi:10.1007/BF00763448

\bibitem{Maartens:1991mj}
R.~Maartens and S.~D.~Maharaj,
``Conformal symmetries of pp waves,''
Class. Quant. Grav. \textbf{8}, 503-514 (1991)
doi:10.1088/0264-9381/8/3/010

\bibitem{Keane:2004dpc}
A.~J.~Keane and B.~O.~J.~Tupper,
``Conformal symmetry classes for pp-wave spacetimes,''
Class. Quant. Grav. \textbf{21}, 2037 (2004)
\href{http://arXiv.org/abs/1308.1683}{{\texttt{arXiv:1308.1683}}}.


\bibitem{Duval:2017els}
C.~Duval, G.~W.~Gibbons, P.~A.~Horvathy and P.~M.~Zhang,
``Carroll symmetry of plane gravitational waves,''
Class. Quant. Grav. \textbf{34}, no.17, 175003 (2017)
\href{http://arXiv.org/abs/1702.08284}{{\texttt{arXiv:1702.08284}}}.

\bibitem{Zhang:2019gdm}
P.~M.~Zhang, M.~Cariglia, M.~Elbistan and P.~A.~Horvathy,
``Scaling and conformal symmetries for plane gravitational waves,''
J. Math. Phys. \textbf{61}, no.2, 022502 (2020)
\href{http://arXiv.org/abs/1905.08661}{{\texttt{arXiv:1905.08661}}}.



%%%%%%%%%%%%%%%%%%%%%% KILLING TENSORS

\bibitem{Keane:2010hg}
A.~J.~Keane and B.~O.~J.~Tupper,
``Killing tensors in pp-wave spacetimes,''
Class. Quant. Grav. \textbf{27}, 245011 (2010)
\href{http://arXiv.org/abs/1011.6401}{{\texttt{arXiv:1011.6401}}}.

\bibitem{Koutras}
A. Koutras and J.E.F. Skea, Computer Physics Communications, 115, 350 (1998).

\bibitem{Rani:2003br}
R.~Rani, S.~B.~Edgar and A.~Barnes,
``Killing tensors and conformal Killing tensors from conformal Killing vectors,''
Class. Quant. Grav. \textbf{20}, 1929-1942 (2003)
\href{http://arXiv.org/abs/0301059}{{\texttt{arXiv:0301059}}}.

\bibitem{Zhao:2024xzo}
Q.~L.~Zhao, P.~M.~Zhang and P.~A.~Horvathy,
``Conformally related vacuum gravitational waves, and their symmetries,''
\href{http://arXiv.org/abs/2403.02230}{{\texttt{arXiv:2403.02230}}}.





%%%%%%%%%%%%%%%%%%%% FERMI COORDINATES 

\bibitem{Manasse:1963zz}
F.~K.~Manasse and C.~W.~Misner,
``Fermi Normal Coordinates and Some Basic Concepts in Differential Geometry,''
J. Math. Phys. \textbf{4}, 735-745 (1963)
doi:10.1063/1.1724316

\bibitem{Marzlin:1994ia}
K.~P.~Marzlin,
``Fermi coordinates for weak gravitational fields,''
Phys. Rev. D \textbf{50} (1994), 888-891
\href{http://arXiv.org/abs/9403044}{{\texttt{arXiv:9403044}}}.

\bibitem{Marzlin:1994wc}
K.~P.~Marzlin,
``On the physical meaning of Fermi coordinates,''
Gen. Rel. Grav. \textbf{26} (1994), 619
\href{http://arXiv.org/abs/9402010}{{\texttt{arXiv:9402010}}}.

\bibitem{Delva:2011abw}
P.~Delva and M.~C.~Angonin,
``Extended Fermi coordinates,''
Gen. Rel. Grav. \textbf{44}, 1-19 (2012)
\href{http://arXiv.org/abs/0901.4465}{{\texttt{arXiv:0901.4465}}}.

\bibitem{Guedens:2012sz}
R.~Guedens,
``Locally inertial null normal coordinates,''
Class. Quant. Grav. \textbf{29} (2012), 145002
\href{http://arXiv.org/abs/1201.0542}{{\texttt{arXiv:1201.0542}}}.

\bibitem{Baskaran:2003bx} 
D.~Baskaran and L.~P.~Grishchuk,
``Components of the gravitational force in the field of a gravitational wave,''
Class. Quant. Grav. \textbf{21}, 4041-4062 (2004)
\href{http://arXiv.org/abs/0309058}{{\texttt{arXiv:0309058}}}.

\bibitem{Rakhmanov:2004eh} 
M.~Rakhmanov,
``Response of test masses to gravitational waves in the local Lorentz gauge,''
Phys. Rev. D \textbf{71}, 084003 (2005)
\href{http://arXiv.org/abs/0406009}{{\texttt{arXiv:0406009}}}.

\bibitem{Delva:2006qa}
P.~Delva, M.~C.~Angonin and P.~T.~Proxy,
``A Comparison between matter wave and light wave interferometers for the detection of gravitational waves,''
Phys. Lett. A \textbf{357}, 249-254 (2006)
\href{http://arXiv.org/abs/0609075}{{\texttt{arXiv:0609075}}}.

\bibitem{Rakhmanov:2008is}
M.~Rakhmanov, J.~D.~Romano and J.~T.~Whelan,
``High-frequency corrections to the detector response and their effect on searches for gravitational waves,''
Class. Quant. Grav. \textbf{25}, 184017 (2008)
\href{http://arXiv.org/abs/0808.3805}{{\texttt{arXiv:0808.3805}}}.

\bibitem{Rakhmanov:2009zz}
M.~Rakhmanov,
``On the round-trip time for a photon propagating in the field of a plane gravitational wave,''
Class. Quant. Grav. \textbf{26}, 155010 (2009)
\href{http://arXiv.org/abs/1407.5376}{{\texttt{arXiv:1407.5376}}}.

\bibitem{Rakhmanov:2014noa}
M.~Rakhmanov,
``Fermi-normal, optical, and wave-synchronous coordinates for spacetime with a plane gravitational wave,''
Class. Quant. Grav. \textbf{31}, 085006 (2014)
\href{http://arXiv.org/abs/1409.4648}{{\texttt{arXiv:1409.4648}}}.


\bibitem{Blau:2006ar}
M.~Blau, D.~Frank and S.~Weiss,
``Fermi coordinates and Penrose limits,''
Class. Quant. Grav. \textbf{23}, 3993-4010 (2006)
\href{http://arXiv.org/abs/0603109}{{\texttt{arXiv:0603109}}}.

%\bibitem{Ashby:1975af}
%N.~Ashby and J.~Dreitlein,
%``Gravitational Wave Reception by a Sphere,''
%%Phys. Rev. D \textbf{12}, 336-349 (1975)
%doi:10.1103/PhysRevD.12.336







%%%%%%%%%%%% MEMORY EFFECTS AND MODEL FOR MERGER ESTIMATES



\bibitem{Favata:2010zu}
M.~Favata,
``The gravitational-wave memory effect,''
Class. Quant. Grav. \textbf{27}, 084036 (2010)
\href{http://arXiv.org/abs/1003.3486}{{\texttt{arXiv:1003.3486}}}.

\bibitem{Favata:2011qi}
M.~Favata,
``The Gravitational-wave memory from eccentric binaries,''
Phys. Rev. D \textbf{84}, 124013 (2011)
\href{http://arXiv.org/abs/1108.3121}{{\texttt{arXiv:1108.3121}}}.


\bibitem{Hait:2022ukn}
A.~Hait, S.~Mohanty and S.~Prakash,
''Frequency space derivation of linear and non-linear memory gravitational wave signals from eccentric binary orbits,''
\href{http://arXiv.org/abs/2211.13120}{{\texttt{arXiv:2211.13120}}}.





%%%%%%%%%%%%%%%%%%%%%%%%%%%%%%%%%. COORDINATES SINGULARITY


\bibitem{Marolf:2002bx}
D.~Marolf and L.~A.~Pando Zayas,
``On the singularity structure and stability of plane waves,''
JHEP \textbf{01} (2003), 076
\href{http://arXiv.org/abs/0210309}{{\texttt{arXiv:0210309}}}.

\bibitem{Wang:2018iig}
T.~Wang, J.~Fier, B.~Li, G.~L\"u, Z.~Wang, Y.~Wu and A.~Wang,
``Singularities of plane gravitational waves in Einstein\textquoteright{}s general relativity,''
Gen. Rel. Grav. \textbf{52} (2020) no.2, 21
\href{http://arXiv.org/abs/1807.09397}{{\texttt{arXiv:1807.09397}}}.


%%%%%%%%%%%%%%%%%%%%%%%%%%%%%%%%%% A GARDER

\bibitem{Tahura:2020vsa}
S.~Tahura, D.~A.~Nichols, A.~Saffer, L.~C.~Stein and K.~Yagi,
``Brans-Dicke theory in Bondi-Sachs form: Asymptotically flat solutions, asymptotic symmetries and gravitational-wave memory effects,''
Phys. Rev. D \textbf{103} (2021) no.10, 104026
\href{http://arXiv.org/abs/2007.13799}{{\texttt{arXiv:2007.13799}}}.

\bibitem{Tahura:2021hbk}
S.~Tahura, D.~A.~Nichols and K.~Yagi,
``Gravitational-wave memory effects in Brans-Dicke theory: Waveforms and effects in the post-Newtonian approximation,''
Phys. Rev. D \textbf{104} (2021) no.10, 104010
\href{http://arXiv.org/abs/2107.02208}{{\texttt{arXiv:2107.02208}}}.

\bibitem{Godazgar:2022pbx}
M.~Godazgar, G.~Macaulay, G.~Long and A.~Seraj,
``Gravitational memory effects and higher derivative actions,''
JHEP \textbf{09} (2022), 150
\href{http://arXiv.org/abs/2206.12339}{{\texttt{arXiv:2206.12339}}}.

\bibitem{Seraj:2021qja}
A.~Seraj,
``Gravitational breathing memory and dual symmetries,''
JHEP \textbf{05} (2021), 283
\href{http://arXiv.org/abs/2103.12185}{{\texttt{arXiv:2103.12185}}}.

\bibitem{BenAchour:2024zzk}
J.~Ben Achour, M.~A.~Gorji and H.~Roussille,
``Nonlinear gravitational waves in Horndeski gravity: scalar pulse and memories,''
JCAP \textbf{05} (2024), 026
\href{http://arXiv.org/abs/2401.05099}{{\texttt{arXiv:2401.05099}}}.

\bibitem{BenAchour:2024tqt}
J.~Ben Achour, M.~A.~Gorji and H.~Roussille,
``Disformal gravitational waves,''
\href{http://arXiv.org/abs/2402.01487}{{\texttt{arXiv:2402.01487}}}.

\bibitem{Heisenberg:2023prj}
L.~Heisenberg, N.~Yunes and J.~Zosso,
``Gravitational wave memory beyond general relativity,''
Phys. Rev. D \textbf{108} (2023) no.2, 024010
\href{http://arXiv.org/abs/2303.02021}{{\texttt{arXiv:2303.02021}}}.

\bibitem{Siddhant:2020gkn}
S.~Siddhant, I.~Chakraborty and S.~Kar,
``Kundt geometries and memory effects in the Brans-Dicke theory of gravity,''
Eur. Phys. J. C \textbf{81} (2021) no.4, 350
\href{http://arXiv.org/abs/2011.12368}{{\texttt{arXiv:2011.12368}}}.



\end{thebibliography}
\end{document}